\documentstyle[proceedings,longtable,numreferences,fancybox,rotating,graphics,amssymb,epsfig]{crckapb}
\rotdriver{dvips}
\newcommand{\bit}{\begin{itemize}}
\newcommand{\eit}{\end{itemize}}
\newcommand{\benu}{\begin{enumerate}}
\newcommand{\eenu}{\end{enumerate}}
\newcommand{\bc}{\begin{center}}
\newcommand{\ec}{\end{center}}
\newcommand{\be}{\begin{equation}}
\newcommand{\ee}{\end{equation}}
\newcommand{\bea}{\begin{eqnarray}}
\newcommand{\eea}{\end{eqnarray}}

\def\ve{\varepsilon}

\def\fm{\hbox{$.\!\!^{\rm m}$}}
\def\degr{\hbox{$^\circ$}}

\def\la{\mathrel{\mathchoice {\vcenter{\offinterlineskip\halign{\hfil
$\displaystyle##$\hfil\cr<\cr\sim\cr}}}
{\vcenter{\offinterlineskip\halign{\hfil$\textstyle##$\hfil\cr
<\cr\sim\cr}}}
{\vcenter{\offinterlineskip\halign{\hfil$\scriptstyle##$\hfil\cr
<\cr\sim\cr}}}
{\vcenter{\offinterlineskip\halign{\hfil$\scriptscriptstyle##$\hfil\cr
<\cr\sim\cr}}}}}
\def\ga{\mathrel{\mathchoice {\vcenter{\offinterlineskip\halign{\hfil
$\displaystyle##$\hfil\cr>\cr\sim\cr}}}
{\vcenter{\offinterlineskip\halign{\hfil$\textstyle##$\hfil\cr
>\cr\sim\cr}}}
{\vcenter{\offinterlineskip\halign{\hfil$\scriptstyle##$\hfil\cr
>\cr\sim\cr}}}
{\vcenter{\offinterlineskip\halign{\hfil$\scriptscriptstyle##$\hfil\cr
>\cr\sim\cr}}}}}
\def\degr{\hbox{$^\circ$}}

\def\utw{\smash{\rlap{\lower5pt\hbox{$\sim$}}}}
\def\udtw{\smash{\rlap{\lower6pt\hbox{$\approx$}}}}

\def\fm{\hbox{$.\!\!^{\rm m}$}}

\def\cs{circumstellar }

\def\is{interstellar }
\def\mkm{{\mu}\rm{m}}

\def\bea{\begin{eqnarray}}
\def\eea{\end{eqnarray}}
\begin{opening}
\title{IN THE KITCHEN   OF DUST MODELING}
\author{N.V. VOSHCHINNIKOV}
\institute{Sobolev Astronomical Institute, St.~Petersburg University, \\
      Universitetskii prosp., 28, 198504 St.~Petersburg, Russia, \\
      e-mail: nvv@astro.spbu.ru}
\end{opening}

\runningtitle{IN THE KITCHEN}

\begin{document}

\begin{abstract}
Dust grains have been detected in various astronomical objects.
Interpretation of observations of dusty objects
includes three components:
1) determination of the materials which  can exist in the solid phase
and the measurements or acquisition of their optical constants;
2) selection of the light scattering theory
in order to convert the
optical constants into the optical properties of particles; and
3) the proper choice of the object model
which includes, in particular, the correct treatment of the radiative
transfer effects.
The current state of the components of dust modeling
and the reliability of  information obtained on
the cosmic dust from transmitted, scattered and emitted radiation
are discussed.
\end{abstract}

\section{Introduction}
\subsection{Observations}
Dust grains have been detected in almost all astronomical
objects from the local  environment of the Earth to very distant galaxies
and quasars. The interaction of radiation with grains
includes two main processes:
dust grains {\it scatter}  and {\it absorb} radiation.
The scattered radiation has the same wavelength as the incident one and
can propagate in any direction.
The radiation absorbed by a grain is transformed into its thermal energy
and the particle {\it emits} at wavelengths
usually longer than the absorbed radiation.
Both processes contribute to {\it extinction} when
the radiation from celestial bodies is attenuated by the
foreground dust in the line of sight, i.e. \\
\centerline{{\it Extinction = scattering + absorption.}}

In general, it is possible to investigate the processes of
{\rm extinction},
{\rm scattering} and
{\rm emission} of radiation by cosmic dust.
{\rm Extinction} is observed when the
light is scattered at the scattering angle $\Theta=0^\circ$
(forward-trans\-mit\-ted radiation).
Corresponding observational phenomena are interstellar extinction and polarization.
In the case when the {\rm scattering} dominates,
an observer sees the radiation
scattered at different angles from
$\Theta=0^\circ$ to $\Theta=180^\circ$.
The scattered radiation comes from the comets, zodiacal light,
reflection nebulae, circumstellar shells, galaxies.
Dust {\rm emission}  occurs in the
H\,II regions, circumstellar shells, interstellar clouds, galaxies, etc.

\subsection{Modeling}

Interpretation of observations of dusty objects
can be divided into
three  steps.  By analogy with cooking, 
the ``kitchen of dust modeling" would outline three
primary factors to prepare this dish:
\begin{enumerate}
\item laying-in {\it provision}

The primary
task is to find  elements which can be converted into the solid species
in the circumstellar/interstellar conditions and to determine the resulting
{{\it materials}}.
The next task is to find or to measure
{\it optical constants}
(refractive indices) of the materials under consideration.

\item choice of {\it equipment}

Selection of
{\it light scattering theory} (``equipment") is an  essential
aspect in  dust modeling.
The chosen method must provide the possibility
of reproducing the most significant features of the observational phenomenon
and to work rather fast in order to give  results in a reasonable time.

\item {\it cooking}

This most important part of the procedure is related to the skill of the cook
(modeler) and includes not only a selection of provision and equipment
but also a proper choice of the method of cooking ---
{\it object modeling}.
\end{enumerate}
Lastly, one needs  to taste the prepared dish, i.e. to compare
the model with observations.
The latter are performed with a limited accuracy which imposes
a corresponding limitation on the claims of the model. From another 
point of view,
a very complicated and detailed model with many parameters is
ambiguous in principle. Further complicating the model,
one should not forget about the principle of
optical equivalence introduced by George Gabriel Stokes 150 years ago:

\begin{center}
\begin{minipage}[c]{10.3cm}
\it It is \underline{impossible} to distinguish two beams which are
the sum of non-coherent simple waves if they have the same
Stokes parameters.
\end{minipage}
\end{center}

So, a judicious restriction on the detailed elaboration of
different components in dust modeling should be found.

\section{Abundances and optical constants}
\subsection{Element abundances and depletions}\label{depl}

Ultraviolet (UV) and optical absorption-line studies
have shown that the \is (gas-phase)
abundances of many elements are lower than
{\it cosmic (reference) abundances}.
The rest of the elements is assumed to be locked in solid particles.
The depletion of an element X is defined by
\begin{equation}
D_{\rm X} = \left. \left[\frac{\rm X}{\rm H}\right]_{\rm g} \right/
            \left[\frac{\rm X}{\rm H}\right]_{\rm cosmic}, \label{dx}
\end{equation}
where [X/H] is an abundance of element X
relative to that of hydrogen.
Here X (or $N_{\rm X}$) and
H (or $N_{\rm H}=N_{{\rm HI}}+2N_{\rm H_2}$) are the column densities of an element X and hydrogen
in a given direction.
The abundances by number are usually expressed
as the number of X atoms per $10^6$ hydrogen nuclei (particles per million,
ppm, hereafter).

For a long time, the reference abundances
were assumed to be equal to solar abundances.
However, Snow and Witt~\cite{sw96}
found that many species in cluster and field B stars, and in young F and G stars
were significantly underabundant (by a factor of 1.5--2.0)
relative to the Sun.
The old (solar) and new (stellar)
abundances for the five main elements forming
cosmic dust grains are as follows:
C~(398ppm/214ppm),
O~(851ppm/457ppm),
Mg~(38ppm/25ppm),
Si~(35.5ppm/18.6ppm) and
Fe~(46.8ppm/27ppm).
New abundances limited the number of atoms incorporated into dust particles.
For example, the dust-phase abundances
in the line of sight to the star  $\zeta$~Oph are
79 ppm for C,
126 ppm for O,
23 ppm for Mg,
17 ppm for Si and
27 ppm for Fe.
The most critical situation occurs for  carbon which is the main component
of many dust models.
The inability to explain the observed \is extinction
using the amount of carbon available in the solid phase resulted in so
called ``{\it carbon crisis}" which has not been resolved up to now.


\subsection{Refractive indices and their mixing}
The complex refractive indices ($m$) or  dielectric functions ($\varepsilon$)
of solids are called {\it optical constants}.
The refractive index is written in the form
$m= n(1+ \varkappa i)$ or $m= n + ki$, where $k=n \varkappa \geq 0$.
The sign of the imaginary part of the refractive index is opposite
to that of the time-dependent multiplier in the presentation of
fields. Note that in the book of van de Hulst~\cite{vdh57}
the refractive index is chosen as $m= n - ki$ whereas
in the book of Bohren and Huffman~\cite{bh83} as $m= n + ki$.

The physical sense of $n$ and $k$ becomes clear
if one considers the solution to the wave equations
in an absorbing medium which is performed in the Cartesian coordinate
system $xyz$.
For an electric field propagating,
for example, in the $z$-direction, we have

\be
\displaystyle\vec E = \vec E_0 \exp \left(-\frac{\omega}{c} kz\right)
\exp \left[-i \omega \left(t -\frac{nz}{c}\right)\right],\label{wave}
\ee
where $\omega = 2 \pi \nu= 2 \pi c / \lambda$ is the circular frequency,
$c$ the speed of light, $t$ time.
As seen from Eq.~(\ref{wave}), the imaginary part $k$
(often named the {\it extinction coefficient or index})
characterizes  damping or absorption  of the wave.
The real part $n$ (the {\it refraction index})
determines the phase velocity of the wave in the medium,
$v_{\rm phase} = c/n$.

The real and imaginary parts of the optical constants are
not independent and may be calculated one from another
using the Kramers--Kronig relations.
When applied to $n$ and $k$, the relations are
\be
n(\omega) = 1 + \frac{2}{\pi} {\mathcal P} \int_{0}^{\infty}
\frac{\Omega k(\Omega)}{\Omega^{2} - \omega^{2}} \, {\rm d}\Omega,
\label{nom}\ee
\be
k(\omega) = - \frac{2\omega}{\pi} {\mathcal P} \int_{0}^{\infty}
\frac{n(\Omega)}{\Omega^2 - \omega^2} \, {\rm d}\Omega,
\label{kom}\ee
where $\cal{P}$ denotes the principal part of the integral.
Equations~(\ref{nom}) and (\ref{kom}) allow one to make some conclusions
on the behavior of the optical constants in different wavelength ranges.
In particular, it is impossible to have a material with
$k=0$ at all wavelengths because in this case the radiation does not
interact with the material ($n=1$ everywhere).

\begin{figure}  \bc
\resizebox{\hsize}{!}{\includegraphics{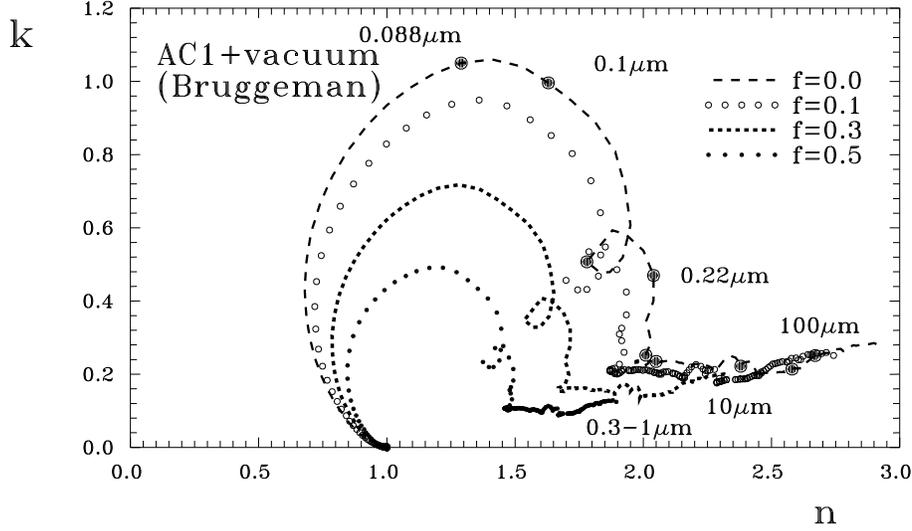}}
\caption{Refractive indices of amorphous carbon calculated
with the Bruggeman mixing rule for  a different fraction of vacuum ($f$=0--0.5).
The values of wavelengths  in $\mkm$ are indicated.}\label{f_riw}
\ec \end{figure}

The optical constants for amorphous carbon plotted in Fig.~\ref{f_riw}
allow us to understand the relation between the real and
imaginary parts of the refractive indices.
In particular,
the limiting and asymptotic values of $n$ and $k$ are clearly seen.
Note that the absorption features  appear as loops in the
$k-n$ diagrams. In the last few years, special measurements of the optical
constants for cosmic dust analogues have been made.
These data are being collected in the database of optical constants for astronomy
(Jena--Petersburg Database of Optical Constants, JPDOC;
see  \cite{heal99},  \cite{ieal02} for details).

The conditions in which cosmic dust grains originate,  grow and evolve
should lead to the formation of heterogeneous particles with complicated structure.
The problem of electromagnetic scattering by such composite
particles is so difficult that a practical, real-time solution is 
currently unfathomable, in particular keeping in mind the  unknown
real structure of grains.
Therefore, the thought of obtaining the optical
properties of heterogeneous particles using
homogeneous particles with  {\it effective}
dielectric functions $\varepsilon_{\rm eff}$
found from a mixing rule (generally called an Effective Medium Theory; EMT)
is an extremely attractive proposition.

Many different mixing rules exist (see
\cite{cval00} and \cite{sih99} for a review).
They are rediscovered from time to time and sometimes can be
obtained one from another.
The most popular EMTs are the classical mixing rules of
(Maxwell) Garnett and
Bruggeman.  In the case of spherical inclusions
and a two-component medium the effective dielectric constant may
be calculated easily from the
dielectric permittivities $\ve_1$, $\ve_2$ and volume fractions
$f$, $1-f$ of the components.
The Garnett rule assumes that one material
is a matrix (host material) in which the other material is embedded.
It is written in the following form:
\be
\displaystyle\ve_{\rm eff} = \ve_2 \left[1 +
\frac{ 3f \frac{\ve_{1} - \ve_{\rm 2}}{\ve_{1} + 2 \ve_{\rm 2}}}
{ 1 - f \frac{\ve_{1} - \ve_{\rm 2}}{\ve_{1} + 2 \ve_{\rm 2}}}
\right].
\ee
When the roles of the inclusion and the host material are reversed,
the inverse Garnett rule is obtained
\be
\displaystyle\ve_{\rm eff} = \ve_1 \left[1 +
\frac{ 3 (1-f) \frac{\ve_{2} - \ve_{\rm 1}}{\ve_{2} + 2 \ve_{\rm 1}}}
{ 1 - (1-f) \frac{\ve_{2} - \ve_{\rm 1}}{\ve_{2} + 2 \ve_{\rm 1}}}
\right].
\ee
The Bruggeman rule is symmetric with respect to the interchange of materials
\be
\displaystyle f \frac{\ve_1 - \ve_{\rm eff}}{\ve_1 + 2 \ve_{\rm eff}} +
(1-f) \frac{\ve_2 - \ve_{\rm eff}}{\ve_2 + 2 \ve_{\rm eff}} =0.
\ee
Figure~\ref{f_riw} shows  variations of
the refractive indices of amorphous carbon calculated
with the Bruggeman mixing rule for  a different fraction of vacuum.

A special question is the range of applicability of the EMTs.
A general conclusion made from calculations (see \cite{vidch98}, \cite{wcg98})
is that
an EMT agrees well with the exact theory if
the inclusions are  Rayleigh and their volume  fraction $f$ is
below $\sim$~40--60\%. In the case of
non-Rayleigh inclusions,  apparently, $f$ should not exceed
$\sim$10\%.
Kolokolova and Gustafson~\cite{kolgust01}
performed a comprehensive study of the possibility of applying nine
mixing rules comparing calculations for organic spheres with
silicate inclusions using microwave analog experiments.
They recommend the use of EMTs if the volume  fraction of inclusions
is not more than 10\%.

\section{Light scattering theories}

When  the optical constants  are chosen,
they can be converted
into {\it optical properties} of particles:
various cross-sections, scattering  matrix, etc.
using a light scattering theory.
Theories of light scattering by particles may make it possible
to calculate the following: \\
1) intensity and polarization of radiation scattered in any direction; \\
2) energy absorbed or emitted by a grain and to find its temperature; \\
3) emission spectra of dusty objects; \\
4) radiation pressure force on  dust grains which frequently
determines their motion.

\enlargethispage{\baselineskip}

Theoretical approaches to solve the light scattering problem
are divided into {\it exact methods} and
{\it approximations} by their dependence on the following: \\
1) the size parameter $x = 2 \pi r/\lambda$, where $r$ is the typical
      size of a particle (e.g., the radius of the equi-volume sphere)
      and $\lambda$ the wavelength of incident radiation
      in the surrounding medium; \\
2) the module of the difference between the refractive index and  unity $|m-1|$; \\
3) the phase shift $\rho = 2x |m-1|$. \\
Approximations may be applicable if at least two of these quantities are
much smaller or much larger than unity \cite{vdh57}.
In particular, approximate approaches allow one to estimate
the wavelength dependence of extinction
cross-sections easily.  The latter
determine, for example,  \is extinction $A(\lambda)$ (see Sect.~\ref{ext_int}).
Multi-color observations
show that $A(\lambda) \propto \lambda^{-1}$ in the visible part of the spectrum and
 no approximation
predicts such a wavelength dependence.
Therefore, astronomers in general are doomed to work with  exact methods.
Spherical grains do not  explain the \is polarization
and another feature required
of the theory is the ability to treat
non-spherical particles with sizes close to or larger than the
radiation wavelength.

At present, only a few methods satisfy astronomical
demands and three of them are widely used in astrophysical modeling.
These are the separation of variables method for spheroids,
the T-matrix method  for  axially symmetric
particles, and some modifications of the method of momentum
(and first of all the discrete dipole approximation, DDA).

The current state of methods and techniques of calculating light scattering
by non-spherical particles  are described in special issues of
{\it Journal of Quantitative Spectroscopy and Radiative Transfer}
\cite{hov96}, \cite{lum98}, \cite{mth99}, \cite{vfc01},
review papers \cite{jonesar99}, \cite{wriedt98}
and the collective monograph  \cite{mht00}.
A detailed description of different approximations can be found in
\cite{jonesar99}, \cite{koh99}  and \cite{mui02}.

\enlargethispage{\baselineskip}

\section{Objects' models}

In the modeling of dusty objects
one needs to select not only optical constants and a light
scattering approach, but also an appropriate model of the object.
The model of a dusty object includes
an appropriate choice of the spatial distribution of scatterers
(dust grains)  and illuminating sources and  correct
treatment of radiative-transfer effects.

Radiative transfer methods are rather
conservative and making changes or modifications usually require
tremendous efforts.
As a result, the {\it radiative transfer code  available  determines the
final result of modeling,}
but different radiative transfer codes  may give different results.
During its long history, radiative transfer theory passed through
several stages:  analytical, semi-analytical and numerical. 
The problems which can be solved analytically are
very simple ones like the fluxes from a sphere found in the Eddington
approximation assuming grey opacity and the spherical phase function.
Modern observational techniques give the spectral energy distributions,
images and polarization maps  of very complicated objects like
fragmented molecular cloud cores, circumbinary and circumstellar disks.
These data cannot be modeled without very complicated radiative transfer
programs, and consideration of polarization usually requires application
of Monte Carlo methodologies.

Some characteristics of  radiative transfer
programs created during the last  25 years and their applications to
the interpretation of observed data on dusty cosmic objects
are collected in Table~7 in \cite{v02}.
The major part of the work mentions that these are based on two
methods: \\
1) iterative schemes to solve the moment equations of the radiative
transfer equation and \\
2) Monte Carlo  simulation. \\
Standard applications include the following:
\cs  shells and envelopes around early
(pre-main-sequence) and late-type stars
and young stellar objects, reflection nebulae,
\is clouds and globules, diffuse galactic light
and in recent years galaxies and active galactic nuclei.

\section{Interstellar extinction and polarization}
\subsection{Observations}\label{ext_int}

Observational analysis of \is extinction and polarization
is twofold: ``in depth'' and ``in breadth''.
The first involves examination
of the {\it wavelength dependence} and
gives information about the properties  of \is grains.
The second includes the study of the
distribution of dust matter and relates to work on  galactic
structure.


The wavelength dependence of {\it extinction} $A(\lambda)$ is rather well
established in the range from near-infrared (IR) to far-UV.
It has often been
represented using simple analytical formulae.
The most recent analytical fitting of the extinction curves
was obtained in \cite{fp99}.
Observations in the far-UV show that the growth of the
\is extinction continues almost up to the Lyman limit
\cite{sassetal01}.
At wavelengths shorter than 912~\AA, the extinction is
dominated by photo-ionization of atoms, not
by the scattering and absorption by dust.
In the extreme UV ($\lambda$\,100--912\,\AA),
almost all radiation from distant
objects is ``consumed'' by neutral hydrogen and helium
in the close vicinity of the Sun.
In the X-range, photo-ionization of other abundant atoms (C, N, O, etc.)
becomes important.
The resulting wavelength dependence of  extinction in the diffuse \is medium
is plotted in Fig.~\ref{fall_ext}.
It shows  extinction cross-sections $\sigma_{\rm d}(\lambda)$ related to the column density of
H-atoms, which can be found from the following expression (see \cite{v02} for details):
\begin{figure}[htb]\bc
\resizebox{\hsize}{!}{\includegraphics{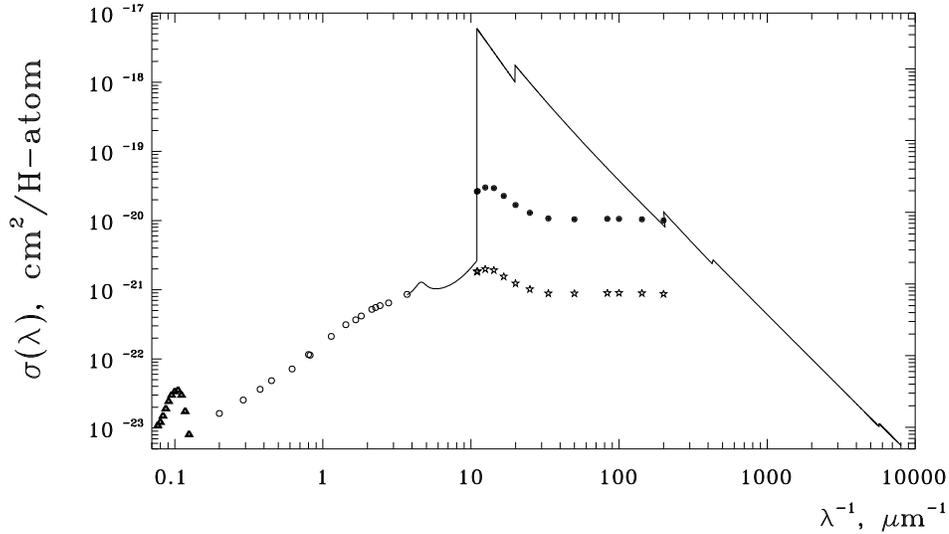}}
\caption{Extinction in the diffuse \is medium. The sources of the data are given in
Table~9 in \protect \cite{v02}.
}\label{fall_ext} \ec \end{figure}
\begin{equation}
\sigma_{\rm d}(\lambda)
= 4.18 \times 10^{-22}\, \frac{A(\lambda)}{A_{\rm V}}. \label{sig1}
\end{equation}
The coefficient in Eq.~(\ref{sig1}) was calculated for the
ratio of the total extinction to the selective one
$R_{\rm V}= A_{\rm V}/E{\rm (B-V)}=3.1$ and
the gas to dust  ratio
${N({\rm H})}/{E{\rm (B-V)}} = 6.83 \times 10^{21} \,
{\rm atoms \, cm^{-2} \, mag^{-1}}$.

\begin{figure}[htb]  \bc
\resizebox{\hsize}{!}{\begin{turn}{90}\includegraphics{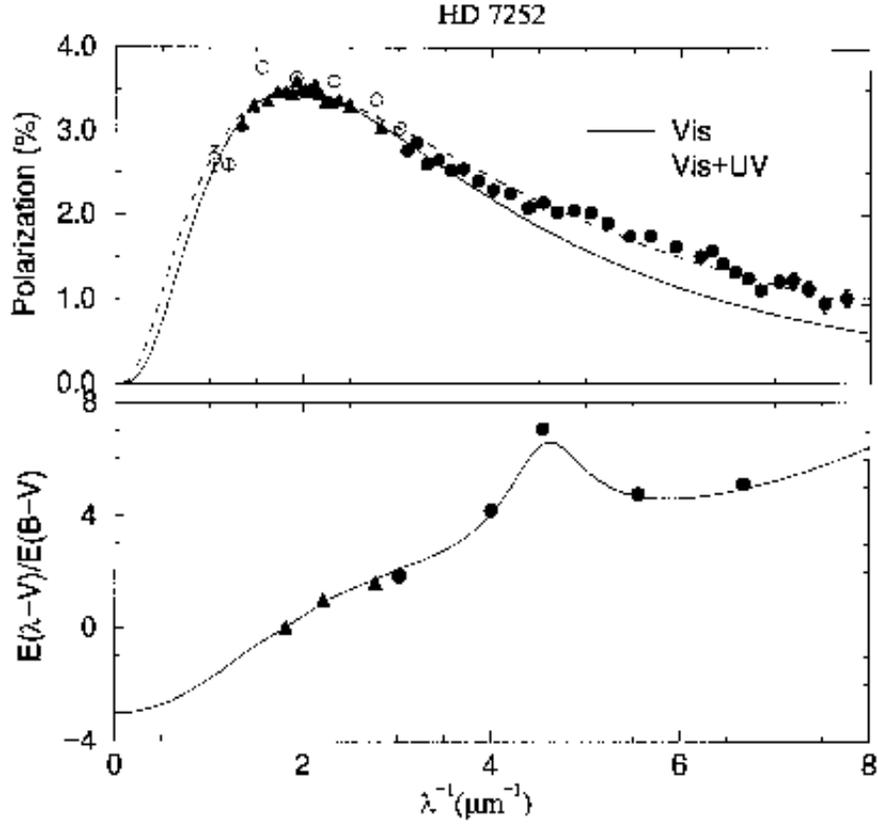}\end{turn}}
\caption{Polarization and extinction curves in the direction of the
star HD~7252 (after \protect \cite{w96b}).}\label{pe}
\ec \end{figure}

The phenomenon of {\it \is linear polarization} is connected to the effect
of the  linear dichroism of the \is medium
which arises because of the  presence of
aligned non-spherical grains.  Such particles produce
different extinctions  of light depending on the  orientation
of the electric vector of incident radiation relative to the particle
axis.
The  wavelength dependence of polarization  $P(\lambda)$
now is known in the spectral range $\lambda$\,0.12--12\,$\mkm$.
The polarization degree usually has a maximum
in visible  and declines in the IR and UV (Fig.~\ref{pe}).
As a rule, the dependence $P(\lambda)$ is  described by
an empirical formula suggested by Serkowski~\cite{serk73}
\be
P(\lambda)/P_{\rm max} = \exp [-K \ln^2 (\lambda_{\rm max}/\lambda)].
\label{serkk}
\ee
Initially, the {\it Serkowski's curve}  had only two parameters:
the maximum degree of polarization $P_{\rm max}$  and the wavelength
corresponding to it $\lambda_{\rm max}$.
The coefficient $K$ was chosen by Serkowski~\cite{serk73} to be equal to 1.15.

An example of the behavior of polarization is shown
in Fig.~\ref{pe}. The solid curves fit the ground-based data
using only  Serkowski's curve (Eq.~(\ref{serkk})) and the relation
$K = 1.86 \cdot \lambda_{\rm max} -0.10$ found in \cite{wilketal82}.
The \is polarization in the direction of
HD~7252 ($\lambda_{\rm max}=0.52\, \mkm$) displays a clear
excess over the extrapolated curve (``super-Serkowski behavior'').
Note that polarization features have been discovered in only
a few directions in the Galaxy, 
although the bump near $\lambda 2175$~\AA~
is a common attribute of all extinction curves.
In order to represent the wavelength dependence of polarization from IR
to UV, Martin {\it et al.}~\cite{mcw99} suggested a five-parameter interpolation
formula consisting of two terms describing UV and visual-IR polarization.
The result of fitting is shown by the dashed curve in Fig.~\ref{pe}.

Note also that
historically the direction of starlight polarization
is associated with the orientation
of the plane-of-the-sky component of the \is
magnetic field, $B_{\bot}$. The data of polarimetric surveys
together with other observations like  Zeeman splitting of the
H\,{\sc I} (21 cm) line are used to study the magnetic field structure at different scales.

\subsection{Interpretation}\label{ip}
\begin{figure}
\bc
\resizebox{\hsize}{!}{\includegraphics{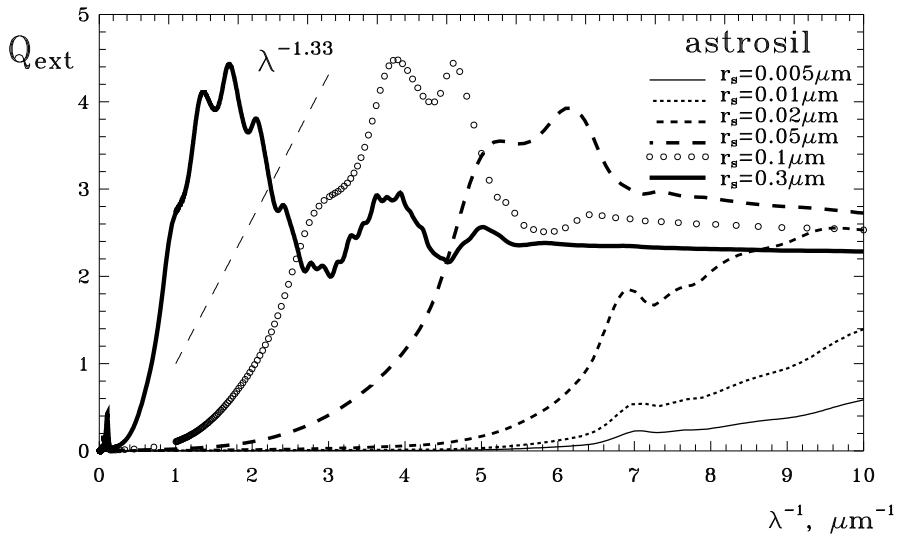}}
\caption{Wavelength dependence of the extinction efficiency factors
for homogeneous spherical particles of different sizes
consisting of astronomical silicate.
The dashed segment shows the approximate wavelength dependence of the mean
galactic extinction curve at optical wavelengths 
(after \protect \cite{v02}).}\label{fq_w}
\ec
\end{figure}

The intensity of radiation after passing through a dust cloud $I(\lambda)$
is equal to
\be
I(\lambda)=I_\star(\lambda)e^{-\tau_{\rm ext}(\lambda)},
\ee
where $I_\star(\lambda)$ is the source (star) intensity and $\tau(\lambda)$
the optical thickness along the line of sight. The \is extinction is
\be
A(\lambda)=-2.5\log \frac{I(\lambda)}{I_\star(\lambda)}\approx 1.086\tau_{\rm ext}(\lambda).
\ee
For spherical particles  of radius $r_{\rm s}$, we have
\begin{equation}
A(\lambda)= 1.086 \,\pi r_{\rm s}^2 \, Q_{\rm ext}(m,r_{\rm s},\lambda) \,N_{\rm d} =
 1.086 \,\pi r_{\rm s}^2 \, Q_{\rm ext}(m,r_{\rm s},\lambda) \,n_{\rm d}\, D,
\label{single_ext}\end{equation}
where $N_{\rm d}$ and $n_{\rm d}$ are the column  and
number densities of dust grains, correspondingly, and $D$ is the distance
to the star.
From Eq.~(\ref{single_ext}) it follows that
the wavelength dependence of \is extinction
is determined completely by the wavelength dependence of the
extinction efficiencies $Q_{\rm ext}$.
Such a dependence is plotted in Fig.~\ref{fq_w} for particles of astronomical
silicate.

\begin{figure}
\bc
\resizebox{\hsize}{!}{\includegraphics{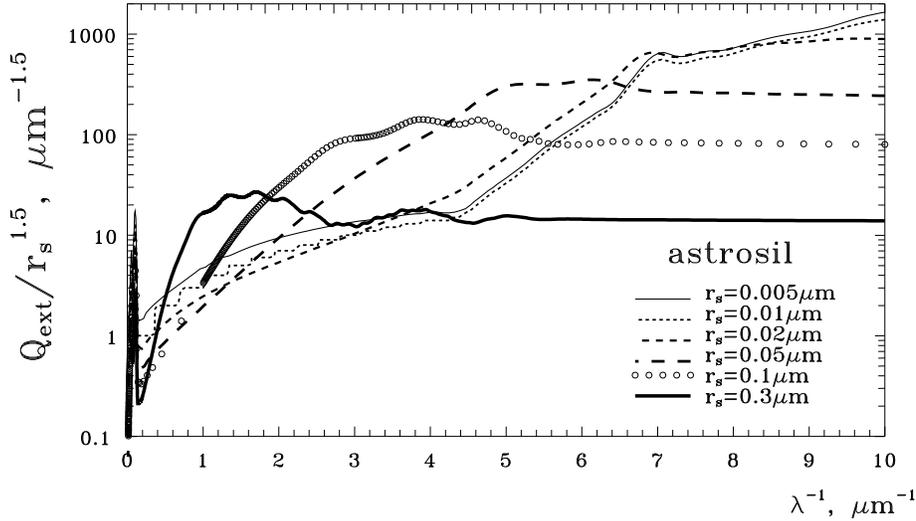}}
\caption{Wavelength dependence of the integrand in Eq.~(\ref{ex})
for homogeneous spherical particles of different sizes
consisting of astronomical silicate.}\label{fq05_w}
\ec
\end{figure}

Dust grains are considered to have some size distribution.
Very often  a grain size distribution
like that suggested in \cite{mrn} (``MRN'')
\mbox{$n(r_{\rm s})\propto r_{\rm s}^{-3.5}$} is used.
In this case, the extinction  is proportional to
\be
A(\lambda) \propto\int Q_{\rm ext}(\lambda)\, r_{\rm s}^{-1.5} {\rm d}r_{\rm s}.
\label{ex}
\ee
Figure~\ref{fq05_w}
allows us to estimate the contribution of
particles of different sizes to the extinction for a given wavelength
(see Eq.~(\ref{ex})) if the distribution is like MRN.

From Fig.~\ref{fq_w}  the influence of the size and
chemical composition of particles on the extinction for a given
wavelength can be examined.
In all cases, the rate of growth increases until we approach the first
maximum of $Q_{\rm ext}(m,x)$.
As follows from Fig.~\ref{fq_w}, spheres of astrosil with
$r_{\rm s} \approx 0.1 \, \mkm$
can produce the dependence $A(\lambda)$ resembling the observed one.
The same is possible for  spheres of amorphous carbon
of smaller radius (see discussion in \cite{v02}).
So, from the wavelength dependence of extinction
one can determine only the product of the typical particle size
on refractive index but not the size or chemical composition of dust grains
separately.
It is also possible to show that particles of different structure
(for example, with mantle or voids) as well as
of different shape may represent
the  dependence of $A(\lambda)$ rather well.
Thus, neither chemical composition, nor structure and shape  of
dust particles can be uniquely deduced from
the wavelength dependence of the \is extinction.

Along with the wavelength dependence, it is important to reproduce the
{\it absolute value of extinction} using the  dust-phase  abundances
found for a
given direction $[{\rm X}/{\rm H}]_{\rm d}$.
These abundances can be expressed via\\
1) observed quantities:
interstellar extinction $A_{\rm V}$ and hydrogen  column density $N({\rm H})$;\\
2) model parameters:
mass of  constituents in a grain,
the relative part of the element $X$ in the constituent $i$,
density of grain material and
relative volume of the constituent in a particle $V_i/V$; and \\
3) a calculated quantity:
the ratio of the extinction cross-section  to the particle volume
$C_{\rm ext}/V$. \\
The last ratio must be maximal in order to produce the
largest observed extinction and simultaneously to save the material.
However, at the moment it is rather difficult
to explain the carbon crisis
using particles of different structure and shape
(see discussion in \cite{v02}).
For example, in order to explain the absolute extinction  of the star $\zeta$~Oph
(HD~149757; $A_{\rm V}=0\fm94$)
using homogeneous spherical particles,
the minimum abundance of carbon must be 320  ppm in the case of amorphous
carbon and 267  ppm in the case of  graphite (the dust-phase value
is 79 ppm, see Sect.~\ref{depl}).
If the particles are of astrosil (MgFeSiO$_4$),
52.5  ppm of Fe, Mg and Si
and 211  ppm of O are required.
All these abundances are larger than those estimated
from observations of dust-phase abundances.
Possibly, the way to resolve the problem
is a thorough analysis of cosmic abundances which probably
are not the same in different galactic regions.


To model the {\it \is polarization} one needs to calculate the forward-transmitted
radiation for an ensemble of  non-spherical aligned dust grains.
This procedure consists of two steps:
1) computing the extinction cross-sections for two polarization modes, and
2)  averaging the cross-sections for given
particle size and orientation distributions.


Let non-polarized stellar radiation passes through a dusty cloud
with a homogeneous magnetic field.
The angle between the line of sight and the magnetic field is
$\Omega$ ($0\degr \leq \Omega \leq 90\degr$).
As follows from observations and
theoretical considerations \cite{dgs79},
the magnetic field determines the  direction of alignment  of dust grains.
The linear polarization produced by a rotating spheroidal particle is
\be
\begin{array}{l}
 \displaystyle P(\lambda) = \int_{\varphi,\,\omega,\,\beta}
 \frac{1}{2} \left[C^{\rm TM}_{\rm ext}(m, r_V, \lambda, a/b, \alpha)
 -C^{\rm TE}_{\rm ext}(...) \right] \times
 \nonumber \\
 \times  \hat{f}[\xi(r_V, \chi'', n_{\rm H}, B, T_d, T_g), \beta] \, \cos 2\psi
\, {\rm d}[\varphi, \omega, \beta] \cdot  N_d  \cdot 100\%,
\end{array}
\label{plam}
\ee
where $C^{\rm TM, \, TE}_{\rm ext}$ are the extinction cross-sections
for two  polarization modes,
$N_{\rm d}$ the column density of dust grains.
$\hat{f}(\xi, \beta)$ is the function of orientation,
depending on the alignment parameter $\xi$
and the precession angle $\beta$.

Of special interest are the
variations of the polarization
in cold dark clouds and star-forming regions.
It was found that in several dark clouds
(L\,1755, L\,1506, B\,216--217
\cite{arceetal98}) the increase of polarization with
growing extinction stops beginning at some value of $A_{\rm V}$.
This fact may be related with the following: \\
1) optics of dust ($C^{\rm TM}_{\rm ext} \simeq C^{\rm TE}_{\rm ext}$
in Eq.~(\ref{plam}));\\
2) physics of dust (no grain alignment);\\
3) object's properties
(line of sight is parallel to the direction of alignment, $\Omega \simeq 0^\circ$
or  two or more clouds are present on the line of sight
which leads to the cancellation of polarization).

\setcounter{figure}{6}
\begin{figure}
 \epsfig{file=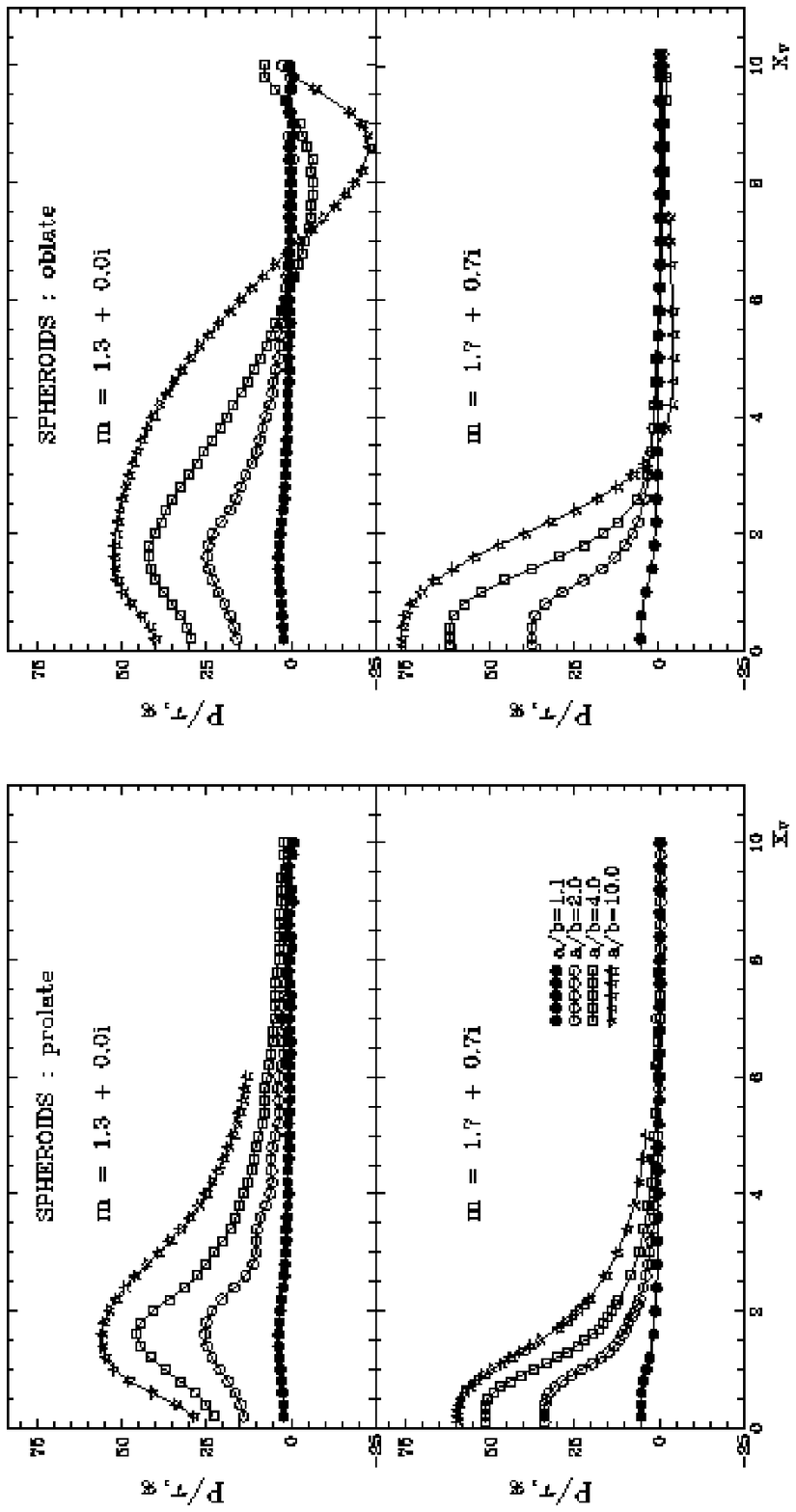,height=18cm,width=9.5cm}
\begin{turn}{90}
{\footnotesize
\setlength{\textheight}{0.5\baselineskip}
\protect\vspace{0.5cm}
{\it Figure \thefigure}. \,\,\,\,
 Polarization efficiency as a function of $x_V$ for
prolate and oblate spheroids with $m=1.3+0.0i$ and $1.7+0.7i$,
picket fence
  }%
\end{turn}
\begin{turn}{90}
{\footnotesize
\setlength{\textheight}{0.5\baselineskip}
\protect\vspace{-0.5cm}
orientation, $\alpha= 90\degr$.
The size/shape effects are illustrated
(after \protect\cite{vihf00}).
  }%
\end{turn}
\label{fabs_f6}
\end{figure}
The proximity of cross-sections
$C^{\rm TM}_{\rm ext}$ and $C^{\rm TE}_{\rm ext}$
may indicate that the shape of grains is more spherical
 in darker parts of clouds.
In a like manner, the reduction of polarization occurs
if the size of non-spherical particles
becomes larger than the radiation wavelength (see Fig.~6).
This Figure illustrates the behavior of the
polarization efficiency $P/\tau$
(the upper/lower sign corresponds to prolate/oblate spheroids):
\be
\frac{P(\lambda)}{\tau(\lambda)} =
\pm \frac{C_{\rm pol}}{C_{\rm ext}} =
\pm \frac{C^{\rm TM}_{\rm ext}-C^{\rm TE}_{\rm ext}}
{C^{\rm TM}_{\rm ext}+C^{\rm TE}_{\rm ext}} \cdot 100\% =
\pm \frac{Q^{\rm TM}_{\rm ext}-Q^{\rm TE}_{\rm ext}}
{Q^{\rm TM}_{\rm ext}+Q^{\rm TE}_{\rm ext}} \cdot 100\%
\label{ppf}
\ee
for non-absorbing and absorbing spheroids
for the case when a maximum polarization is expected
(non-rotating particles, radiation propagates
perpendicular to the symmetry/rotation axis of the spheroid, $\alpha = 90\degr$).
It is seen that  relatively large particles produce
{\it no polarization} independently of their shape.
For absorbing particles,  non-zero polarizations occur at smaller
$x_V$ values\footnote{$x_V = 2 \pi r_V/ \lambda$ where $r_V$ is the radius of
a sphere whose volume is equal to that of the spheroid, $\lambda$ the wavelength
of incident radiation.}
 than for non-absorbing  ones. However, the position at which the ratio
$P/\tau$ reaches a maximum is rather stable in every panel of Fig.~6
and is independent of $a/b$.

\setcounter{figure}{7}
\begin{figure}
\epsfig{file=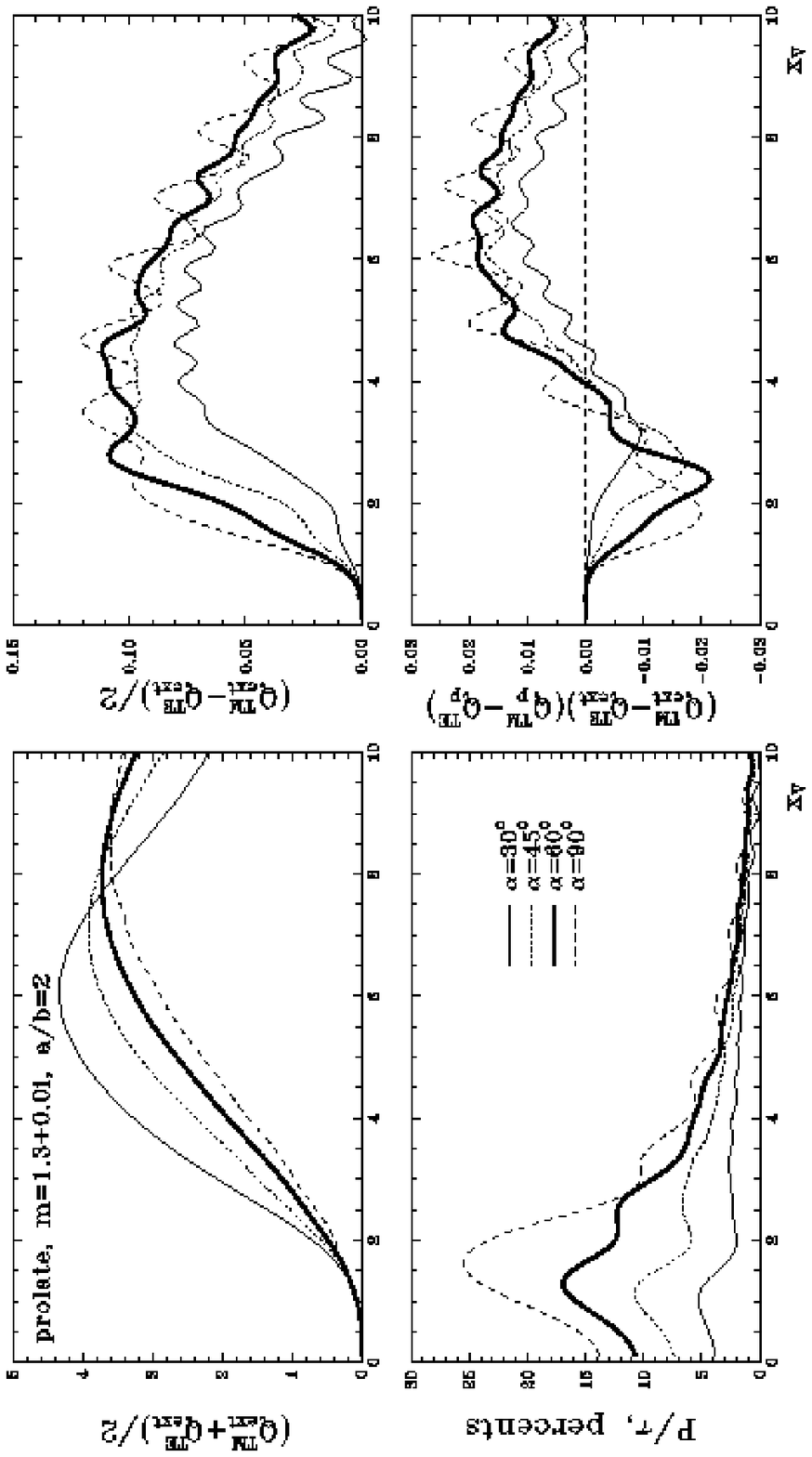,height=18cm,width=9.5cm}
\begin{turn}{90}
{\footnotesize
\setlength{\textheight}{0.5\baselineskip}
\protect\vspace{0.5cm}
{\it Figure \thefigure}. \,\,\,\,
Extinction and linear polarization factors, polarization efficiency
and  circular polarization factors as a function of $x_V$ for
  }%
\end{turn}
\begin{turn}{90}
{\footnotesize
\setlength{\textheight}{0.5\baselineskip}
\protect\vspace{-0.5cm}
prolate spheroids with
$m=1.3+0.0i$ and $a/b=2$, picket fence orientation.
The effect of variations of particle orientation is illustrated
  }%
\end{turn}
\begin{turn}{90}
{\footnotesize
\setlength{\textheight}{0.5\baselineskip}
\protect\vspace{-0.5cm}
(after \protect \cite{v02}).}
\end{turn}
\label{fpol_eff}
\end{figure}
These results no longer remain valid when the radiation is incident obliquely
as is demonstrated in Fig.~7.
The angular dependence of the extinction and linear
polarization factors in Eq.~(\ref{ppf})  differs: if $\alpha$ decreases,
the position of the maximum for $Q_{\rm ext}$
shifts to smaller values of $x_V$ while that for
$Q_{\rm pol}$ shifts to larger $x_V$
(Fig.~7, upper panels).
As a result, the maximum  polarization efficiency for prolate spheroids
is reached at smaller $x_V$ in the case of tilted orientation
(Fig.~7, lower left panel), and the picture is reversed for
oblate particles (see \cite{v02}).

Thus, for particles  larger than the radiation wavelength,
the linear polarization is expected to be rather small. This does not
allow one to distinguish between the particle properties like refractive index,
shape, orientation. Even in the case of the most preferable orientation,
large particles (possibly available in dark clouds)
do not polarize the transmitted radiation significantly.

\section{Scattered radiation}
\subsection{Observations}\label{s_obs}

If a dusty object radiates at a given wavelength,
{\it scattered light} is always present in its radiation.
The amount may be negligible, but it exists.
This is due to the fact that the particle albedo can never be equal
to zero. 

The necessary condition for the appearance of scattered radiation
is the existence of a source of illumination.
The spectrum of the scattered light generally resembles that
of the source. However, the color of the reflection
nebulae may be bluer or redder than that of the illuminating
stars depending on the properties (primarily the size) of the 
dust grains and the scattering geometry
(see \cite{v77} for discussion).
The latter (i.e.,
the mutual position of an illuminating  source, scattering volume
and an observer) also strongly influences the observed
{\it polarization}, which is the usual attribute of the scattered
radiation. The degree of linear polarization
in reflection nebulae may reach 50\% and more.
The typical intrinsic polarization in \cs shells is several percent,
but may sometimes reach 20--30\%.
The centro-symmetric polarization pattern in nebulae
is used to search for the positions of illuminating
stars.

\begin{figure}
\bc
\resizebox{\hsize}{!}{\includegraphics{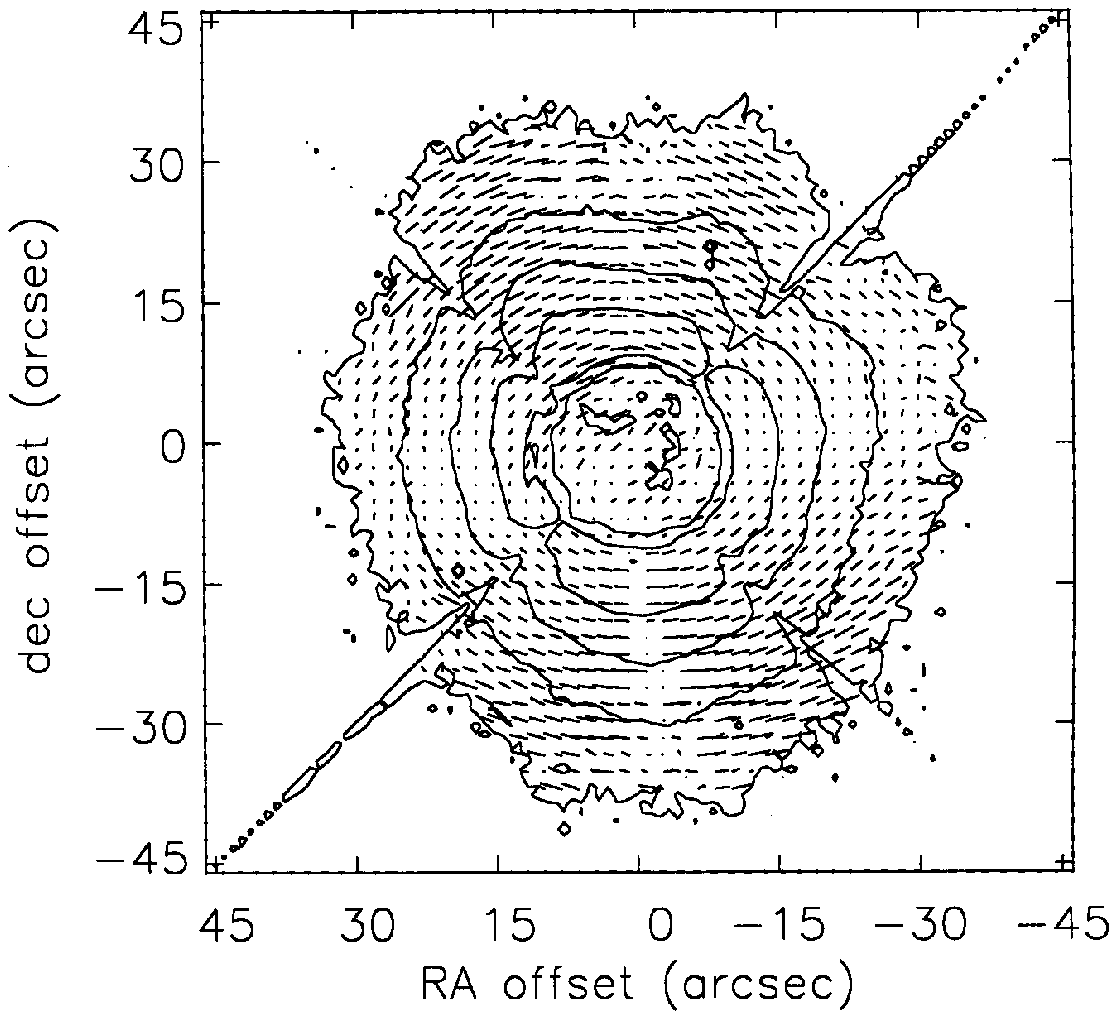}}
\caption{2.2~$\mkm$ polarimetric map constructed from a set of occulted,
polarimetric images of IRC~+10~216. The map is superimposed on a contour
plot of an unpolarized image constructed from the same set of images
(after \protect\cite{kw96}).}
\label{f_irc}
\ec
\end{figure}

However, simple cases,
like  single scattering  from one source,
are not typical in astronomy.
More often, complex or mixed cases are
observed:  attenuation of a part of the scattered
volume by a foreground dust cloud, multiple scattering
effects, etc.  This leads to the appearance
of so-called polarization null points
on the polarization maps
where a reversal of the polarization occurs,
change of the sign of the polarization with wavelength,
non-cent\-ro\-sym\-met\-ric structure of the polarization maps,
variations of the positional angle of polarization with wavelength,
or very high degrees of circular polarization
(see \cite{wvh02} for references and discussion).
Some of these features are seen in Fig.~\ref{f_irc}
where the polarimetric map of the \cs shell around
the carbon star IRC~+10~216 is shown.


\subsection{Interpretation}

The process of light scattering is described in
terms of the particle {\it albedo} and {\it scattering matrix}.
But it is enough to calculate
the first element of the scattering matrix $F_{11}$
({\it phase function}) only
if the polarization is not considered.
Frequently, the phase function is used in the simplified form
suggested by Henyey and Greenstein~\cite{hg41} and is parameterized
by the  {\it asymmetry parameter} $g$ (or $\langle \cos \Theta \rangle$).
It varies from --1 (mirror particles) to 1 (all radiation is
forward  scattered).
Note that mirror particles (with
$g<0$)  are  very atypical
because no grain material (with the exception of very small
iron particles) gives at visual
wavelengths a negative asymmetry parameter.  Even pure conductors
($m= \infty $) have $g \approx -0.2$ \cite{vdh57}.

In optically thin case the amount of observed scattered radiation is
proportional to
\be
I(\lambda) \propto I_\star(\lambda) {[1 - g(\lambda)] \Lambda(\lambda)}
\tau_{\rm ext}(\lambda), \label{sca}
\ee
where  $I_\star(\lambda)$ describes the power of the source,
the quantities ${\Lambda}$ and $1 - g(\lambda)$
characterize the ability of a particle to scatter
the radiation and the scattering geometry, respectively, and
the last multiplier $\tau_{\rm ext}(\lambda)$
is proportional to the extinction cross-section and the number of scatterers.
It is evident that the incoming scattered radiation vanishes
if $I_\star(\lambda), \Lambda(\lambda)$ or
$\tau_{\rm ext}(\lambda)$ approaches zero or
$g(\lambda)$ tends to unity.

\setcounter{figure}{9}
\begin{figure}
 \epsfig{file=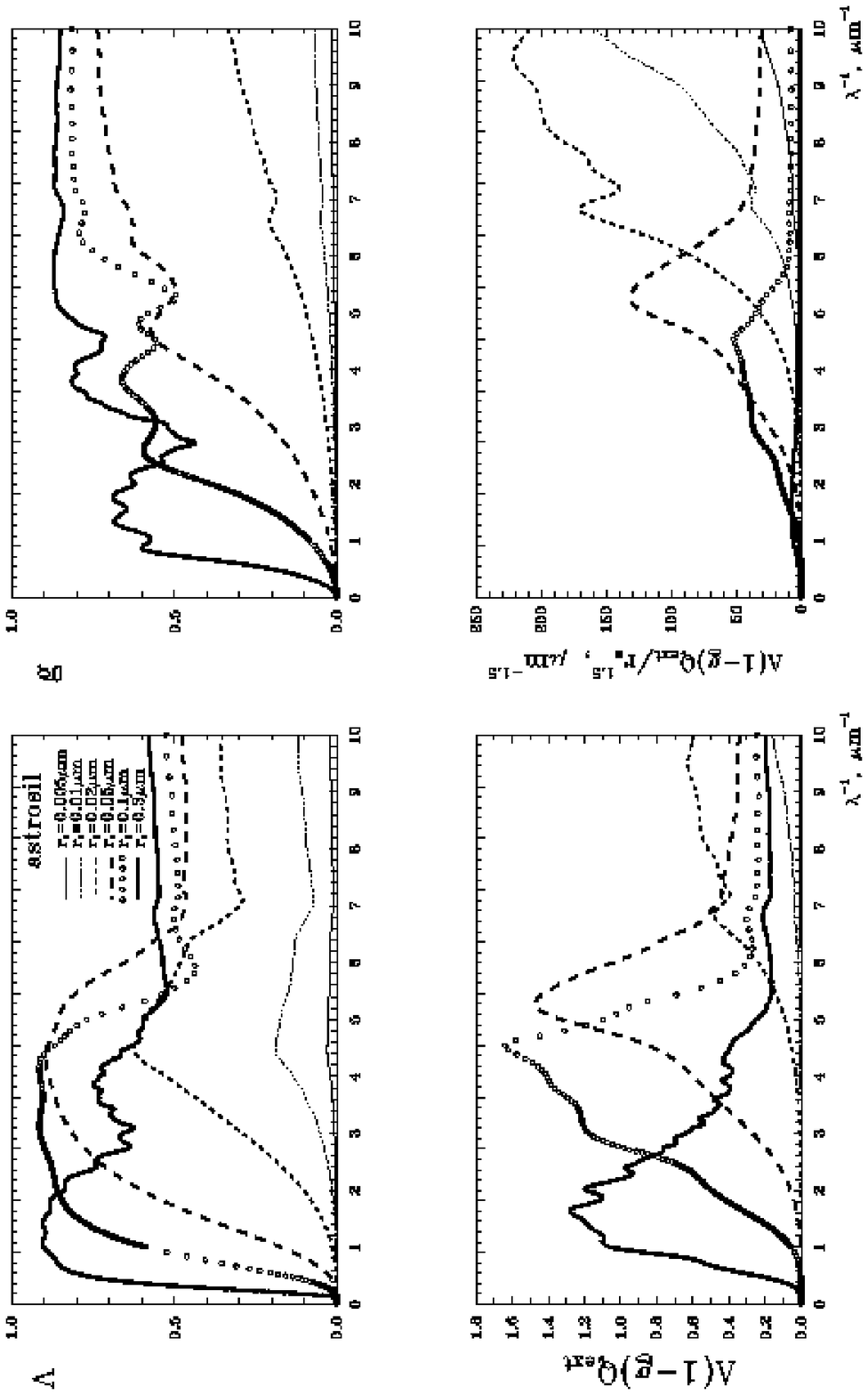,height=18cm,width=9.5cm}
\begin{turn}{90}
{\footnotesize
\setlength{\textheight}{0.5\baselineskip}
\protect\vspace{0.5cm}
{\it Figure \thefigure}. \,\,\,\,
Wavelength dependence of the albedo, asymmetry parameter and
the product of grain scattering characteristics given by
  }%
\end{turn}
\begin{turn}{90}
{\footnotesize
\setlength{\textheight}{0.5\baselineskip}
\protect\vspace{-0.5cm}
Eq.~(\ref{sca})
for homogeneous spherical particles of different sizes
consisting of astronomical silicate.
The lower right panel allows one to
  }%
\end{turn}
\begin{turn}{90}
{\footnotesize
\setlength{\textheight}{0.5\baselineskip}
\protect\vspace{-0.5cm}
estimate the contribution
of particles into scattered radiation
if the size distribution is like MRN.
}%
\end{turn}
\label{f_w}
\end{figure}
The standard behavior of particle albedo and asymmetry parameter
is the following: $\Lambda \ll 1$ and $g \simeq 0$ for small size
parameters (small sizes or large wavelengths),
both characteristics grow with increasing
particle size or decreasing the radiation wavelength
and reach the asymptotic values for very large size
parameters.
The wavelength dependencies of $\Lambda$ and $g$ are
shown in Fig.~9 (upper panels) for particles of astrosil.
It is seen that the albedo of particles with $r_{\rm s} \geq 0.05 \,\mkm$
is rather high (up to $\sim$0.9) in a wide wavelength range and reduces
in the far-UV.
At the same time, the asymmetry parameter shows a tendency toward growth.
So, we expect reduction of the role of scattered light in the UV
in comparison with the visual part of spectrum. This is clearly
seen at the lower left panel of Fig.~9 where
the product $[1 - g(\lambda)] \Lambda(\lambda) Q_{\rm ext}(\lambda)$
as given by Eq.~(\ref{sca}) is plotted.
The contribution of a particle of given size to
the scattering occurs at some wavelength which correlates with
the particle size (i.e. larger particles scatter radiation at longer
wavelengths). But grains with radii $r_{\rm s} \la 0.01 \,\mkm$
hardly affect the scattered radiation. This conclusion remains valid
if one considers size distributions like MRN
(see the lower right panel of Fig.~9).

It is important to keep in mind that the {\it albedo and asymmetry parameter
cannot be determined from observations separately,
but only in the form of a combination}. Therefore,
models with one fixed parameter and the other varying
make little physical sense. The dependence of $\Lambda$ on
$g$ is plotted in Fig.~10 for particles from astrosil and AC1.
It is seen that some pairs of parameters correspond to
{\it no particles}. The theoretical constraints on
the albedo and asymmetry parameter were discussed by Chlewicki
and Greenberg~\cite{cg84} who showed that some modeling results 
could not be represented in the optics of small particles.
\setcounter{figure}{10}
\begin{figure}
 \epsfig{file=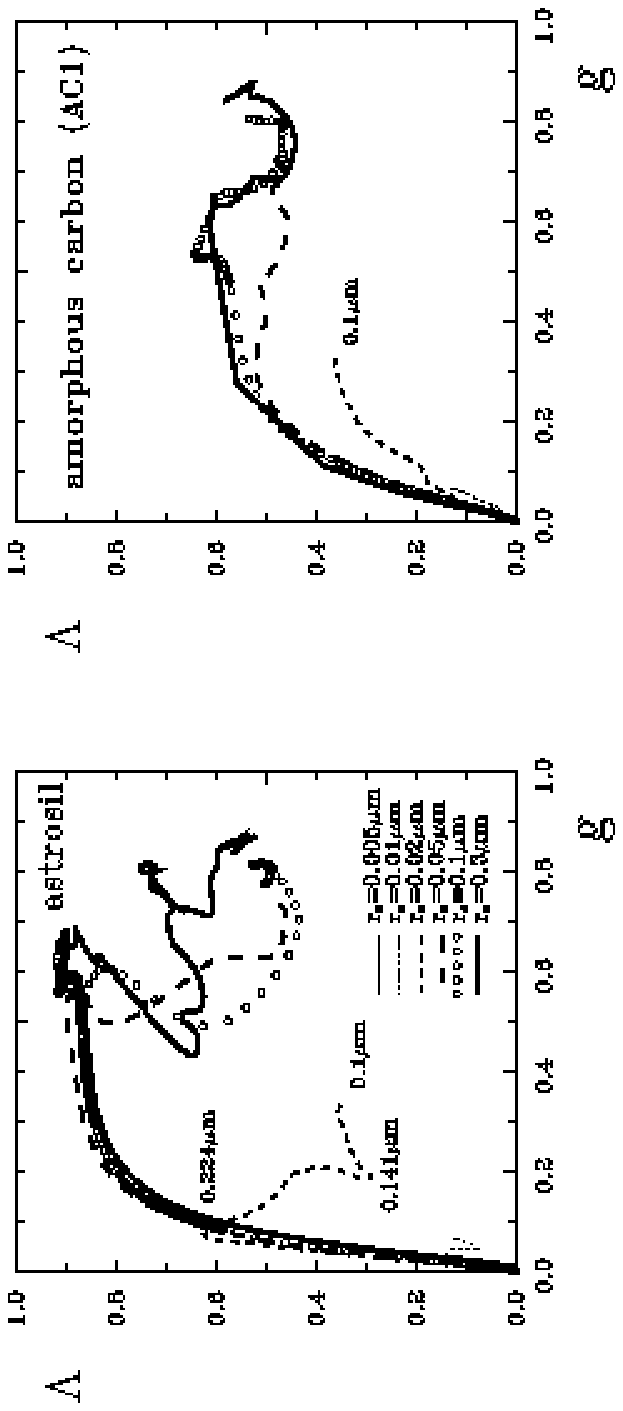,height=18cm,width=9.5cm}
\begin{turn}{90}
{\footnotesize
\setlength{\textheight}{0.5\baselineskip}
\protect\vspace{0.5cm}
{\it Figure \thefigure}. \,\,\,\,
Albedo dependence on asymmetry parameter
for homogeneous spherical particles of different sizes
consisting of 
  }%
\end{turn}
\begin{turn}{90}
{\footnotesize
\setlength{\textheight}{0.5\baselineskip}
\protect\vspace{-0.5cm}
astronomical
silicate and amorphous carbon.
The values of wavelength are given in $\mkm$
for particle with $r_{\rm s}=0.02\,\mkm$.
}%
\end{turn}
\label{a_gw}
\end{figure}

In general, the radiation scattered by aligned non-spherical particles
has an azimuthal asymmetry that provokes a non-coincidence of
the directions of the radiation pressure force
and of the wave-vector of incident radiation
\cite{vi83}.
Another consequence of the azimuthal asymmetry is the anisotropy of
the phase function in the  left/right direction.
The geometry of the phase function
in forward/backward and left/right directions
may be characterized by two asymmetry parameters  $g_{||}$
and $g_{\bot}$, respectively (see \cite{v01} for details).
The values of radial asymmetry factor $g_{||}$ decrease
with a growth of $\alpha$ when the path of radiation reduces
from $2a \ (\alpha=0^{\circ})$ to $2b \ (\alpha=90^{\circ})$.
The transversal asymmetry factor $g_{\bot}$ can be rather
large and even exceeds the radial one.
Because the geometry of light scattering by very elongated spheroids
approaches  that of infinite cylinders,\footnote{In this
case the scattered radiation forms the conical surface
with the opening angle $2 \alpha$.}
{\it such particles scatter more radiation ``to the side"
than in the forward direction.}

At the same time,
the albedo for large non-spherical particles becomes close to
that of spheres.
Our calculations made for  particles with different absorption \cite{vihf00}
demonstrate that the distinction of the albedo for spheres and spheroidal particles
remains rather small (within $\sim 20\,\%$) if the ratio
of the imaginary part of the refractive index to its real part
$k/n \ga 0.2 - 0.3$.

The difference in the single light scattering by a spherical
particle and an aligned non-spherical  particle are readily apparent in
the behavior of the elements of the first column of the
scattering  matrix.
These elements determine the scattered radiation
if the incident radiation is non-polarized.
In contrast to spheres, the scattering by
non-spherical particles 
causes  the  rotation of the  positional angle  of linear polarization
and produces circular polarization after the
first scattering event. This is the result of non-zero elements
$F_{31}$ and $F_{41}$ of the scattering matrix
for aligned non-spherical particles.
Thus, some observational features mentioned in Sect.~\ref{s_obs}
may be attributed to light scattering by non-spherical grains.

\setcounter{figure}{11}
\begin{figure}
 \epsfig{file=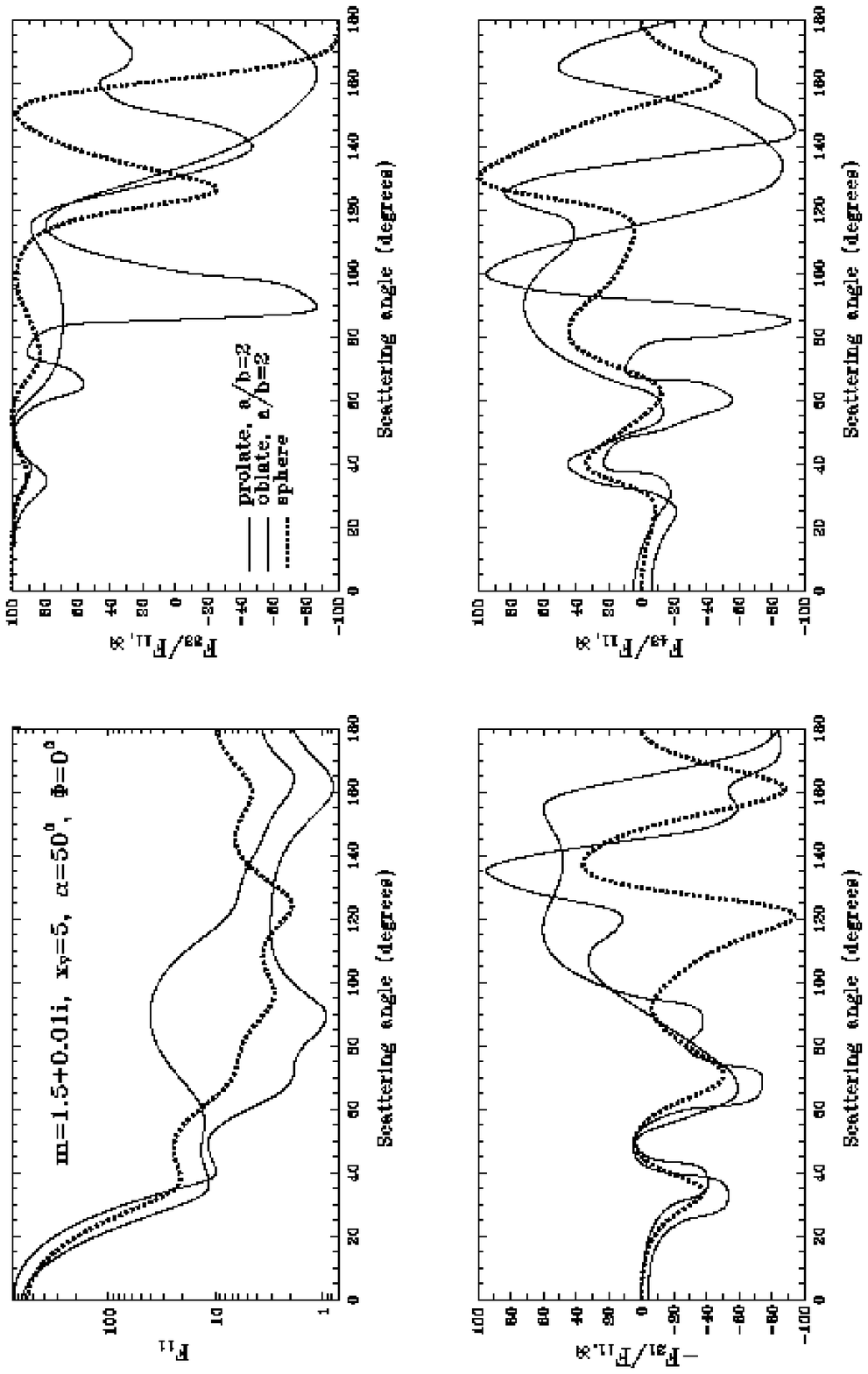,height=18cm,width=9.5cm}
\begin{turn}{90}
{\footnotesize
\setlength{\textheight}{0.5\baselineskip}
\protect\vspace{0.5cm}
{\it Figure \thefigure}. \,\,\,\,
The element $F_{11}$ and the ratios of elements of the scattering matrix
$-F_{21}/F_{11}$, $F_{33}/F_{11}$ and $F_{43}/F_{11}$ for sphere, prolate
}%
\end{turn}
\begin{turn}{90}
{\footnotesize
\setlength{\textheight}{0.5\baselineskip}
\protect\vspace{-0.5cm}
and
oblate spheroids, $m=1.5+0.01i$, $x_V=5$,
$\alpha=50^\circ$, $\Phi=0^\circ$. The influence of variations of particle shape
is illustrated. Adapted 
  }%
\end{turn}
\begin{turn}{90}
{\footnotesize
\setlength{\textheight}{0.5\baselineskip}
\protect\vspace{-0.5cm}
from 
\protect\cite{hovetal96}.
}%
\end{turn}
\label{f_mat}
\end{figure}
The elements of the scattering matrix for spherical and
spheroidal particles of the same size and refractive index are compared in
Fig.~11.
It shows that the major differences
between spheres and spheroids as well as between prolate and
oblate particles appear at large scattering angles.
Such behavior is rather common
(see \cite{mt94a} for discussion).
Therefore, in order to investigate the shape effects,
we need to observe an object located
behind the illuminating source that occurs on occasion.
However, a thin layer of the foreground dust
can produce much more scattered radiation than background dust
(see upper left panel in Fig.~11).
So, it is rather difficult to diagnose the particle shape
from scattered radiation in complex celestial objects.


\section{Infrared radiation}
\subsection{Observations}

\begin{figure}[htb]\bc
\resizebox{\hsize}{!}{\includegraphics{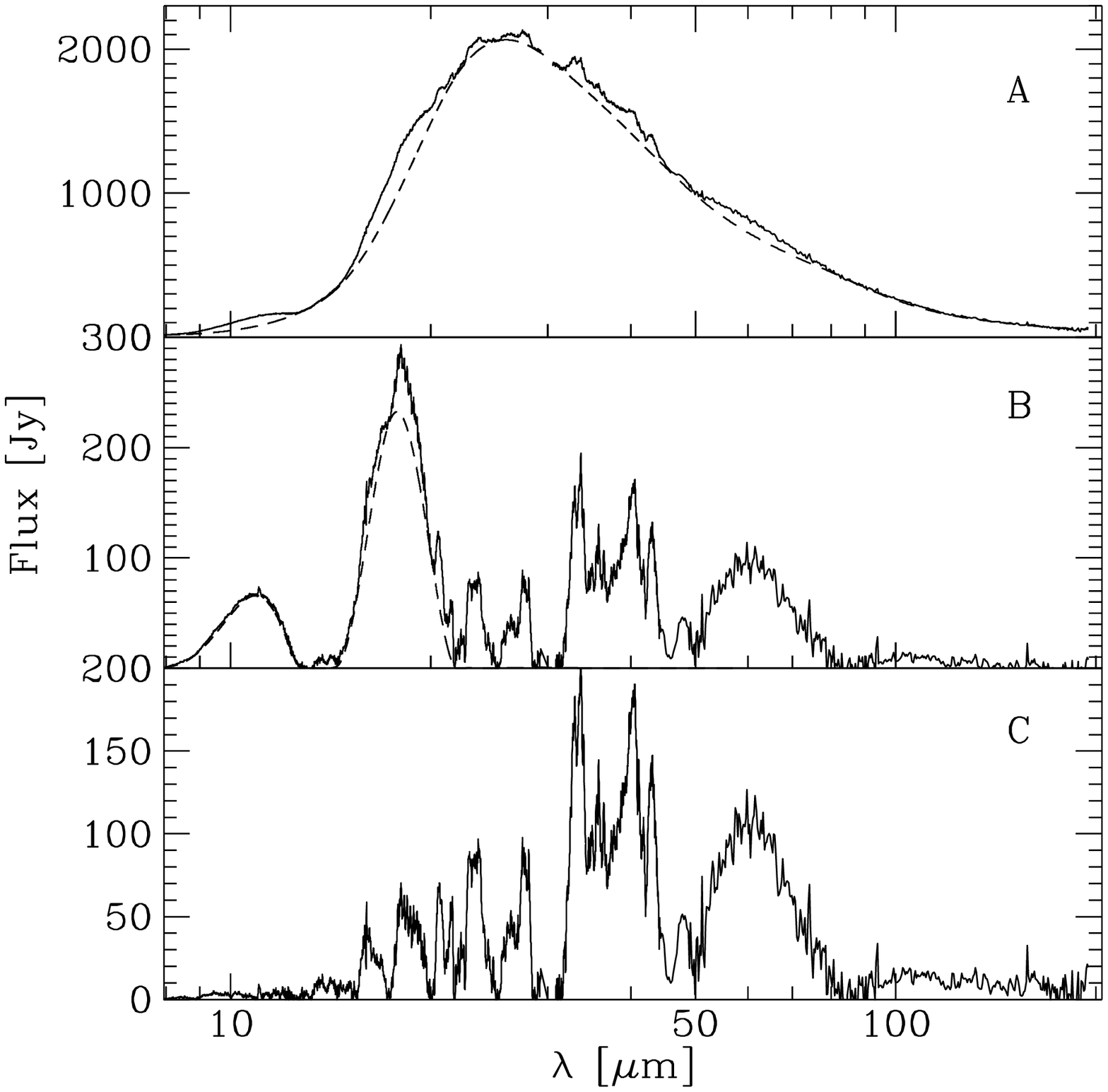}}
\caption{
The ISO spectrum of the post-red supergiant star AFGL~4106 with the spline
fit of continuum. The result of the subtraction of the spline fit continuum
(dashed line in A) is shown in graph B. The result of the removal
of the broad silicate features (dashed line in B) is given in graph C 
(after \protect\cite{metal99}).}\label{gl4106}
\ec
\end{figure}

As noted by Glass~\cite{g99}:
``The central astronomical role of dust is at its most evident in the infrared."
The observed IR and submillimeter emission from interstellar clouds,
circumstellar envelopes, and galaxies is generally thermal emission of
dust heated by stellar radiation or shock waves.
The spectrum of dust emission is blackbody. Its shape is mainly determined by the
properties of heating sources and the dust distribution around them.
The individual characteristics of particles manifest themselves  in
different temperatures of metallic and dielectric grains and in a full
manner as spectral features superimposed on continuum emission.
The dust features are observed in absorption and emission and
are the vibrational transitions
in solid materials --- constituents of grain cores and mantles
(usually bending or stretching modes).
The IR spectral features present the most reliable method of
diagnostics of dust chemical
composition. As a rule, the dust bands  are rather weak and in
order to find them the process of subtracting the continuum
is required (see Fig.~\ref{gl4106}).

The observed IR features attributed to the \is and \cs dust are collected in
Table~\ref{dust_f}. This is the updated version of Table~7 published
in 1986 (see \cite{v86}) which included 24 features and was significantly
enlarged after the discoveries made from the Infrared Space Observatory
(ISO). First of all ISO contributed into the mid-IR where numerous
emission features of crystalline silicates were found.
They were identified with magnesium--iron silicates:
pyroxenes (Mg$_{x}$Fe$_{1-x}$SiO$_3$) and
olivines (Mg$_{2x}$Fe$_{2-2x}$SiO$_4$), where $0\leq x \leq 1$.
In particular, enstatite (MgSiO$_3$)
and forsterite (Mg$_2$SiO$_4$) are the extreme cases
of pyroxenes and olivines with $x=1$, respectively.
Note that the positions and widths of features in Table~\ref{dust_f}
vary from object to object and sometimes a feature was found in one
object only. Several features like the 3.4\,$\mkm$ band
reveal sub-features. Additional information about observational
data can be found in recent publications
\cite{boogetal00},
\cite{bsg99},
\cite{chiar00},
\cite{iso},
\cite{duley00},
\cite{fabaetal99},
\cite{geraetal99},
\cite{gibbetal00},
\cite{gordonetal00},
\cite{m00},
\cite{schuetal98},
\cite{setal02},
\cite{spoonetal02},
\cite{tw97}.

\setlongtables
{\small
\begin{longtable}{ccccc} \\
\caption{Observed IR features attributed to dust}\label{dust_f}\\
\noalign{\smallskip} \hline \noalign{\smallskip}
$\lambda$, {\rm $\mu$m} & $\Delta\lambda^\ast$, {\rm $\mu$m}  & {\rm A/E} & {\rm Identification$^{\ast\ast}$} & {\rm Objects} \\
\noalign{\smallskip}\hline\noalign{\smallskip}
\endfirsthead
\noalign{\smallskip}\hline
\endfoot
\caption{(Continued)}\\
\noalign{\smallskip}\hline\noalign{\smallskip}
$\lambda$, {\rm $\mu$m} & $\Delta\lambda^\ast$, {\rm $\mu$m}  & {\rm A/E} & {\rm Identification$^{\ast\ast}$} & {\rm Objects} \\
\noalign{\smallskip} \hline\noalign{\smallskip}
\endhead
1.15  & 0.2  & E &  silicon nanoparticles?   & 7 \\ 
1.50  & 0.2  & E &  $\beta$-FeSi$_2$?   & 7 \\ 
2.70  & 0.05     & A &  CO$_2$   & 2 \\ 
2.75  &      & A &  phyllosilicates? (O--H)   & 1 \\ 
2.78  & 0.05     & A &  CO$_2$   & 2 \\ 
2.97  &      & A &  NH$_3$ (N--H)         & 2, 9 \\ 
3.07  & 0.7  & A &  H$_2$O (O--H)  &  1, 2, 8, 9 \\ 
3.25  & 0.07 & A &  hydrocarbons (C--H)       & 2    \\
3.28  & 0.05 & A/E &  PAH  (C--H) & 1, 2, 5, 7, 8 \\ 
3.4    & 0.08 & A/E & hydrocarbons, HAC & 1, 2, 5, 7--9 \\ 
3.47  &  0.10--0.15 & A &  hydrocarbons        & 2  \\ 
3.473  &  0.03 & A &  H$_2$CO?, C$_2$H$_6$?        & 2  \\ 
3.5   & 0.08 & E &  H$_2$CO    & 6 \\
3.53  & 0.08  & E & carbonaceous mater. (C--C)   & 6 \\ 
3.53  & 0.025--0.27  & A &  CH$_3$OH (C--H)   & 2 \\ 
3.9   & 0.07 & A &  H$_2$S            & 2 \\
4.1  &  & A & HDO?, SO$_2$?           & 2 \\  
4.27  & 0.02--0.05 & A &  CO$_2$           & 1, 2, 8 \\  
4.38  &       & A  &  $^{13}$CO$_2$          & 1, 2    \\
4.5   &       & A  &  H$_2$O          &  2    \\ 
4.61  & 0.1  & A &  `XCN' (--C$\equiv$N)   & 2, 8, 9 \\  
4.67  & 0.03--0.22 & A & CO     & 2, 8, 9 \\  
4.90   & 0.07 & A &  OCS              & 2, 9 \\
5.25  &      & E &  PAH               & 2, 5, 7, 8 \\ 
5.5--5.6  &  0.06--0.12  & A/E &  metal carbonils (--C=O--)  & 2, 5 \\
5.7   &      & E &  PAH               & 2, 5, 7, 8 \\ 
5.83   &     & A &  HCOOH (--C=O--), H$_2$CO & 2, 9 \\
5.95   &     & A &  carbonaceous mater. (--C=O--)     & 1 \\ 
6.0   & 0.7  & A &  H$_2$O  (O--H)    & 2, 8 \\  
6.14   &      & A &  NH$_3$  (N--H)   & 2 \\ 
6.2   & 0.2  & A/E &  PAH  (--C=C--)  & 1, 2, 5, 7, 8 \\ 
6.8--6.9  & 0.15  & A/E &  HAC, hydrocarbons?  &  2, 5, 8\\  
6.85  & 0.7  & A &  CH$_3$OH          & 2, 9 \\
7.24   & 0.1  & A & HCOOH?          & 2 \\ 
7.25   &     & A & hydrocarbons, HAC       & 1,  2, 8  \\   
7.41   & 0.08  & A & CH$_3$HCO?, HCOO$^-$?      & 2 \\ 
7.67   &  0.06  & A & CH$_4$           & 2, 5, 8 \\   
7.7   & 0.5  & A/E &  CH$_4$, CO      & 2, 5, 7, 8 \\
7.7   &   & E   &  PAH (C--C)     & 1, 2, 5, 7, 8 \\   
8.3    & 0.42  & E &    & 3, 7 \\  
8.7    & 0.36  & E &  PAH (H--C--H)   & 1--5, 7, 8 \\  
9.0   &      & A &  NH$_3$            & 2  \\ 
9.14   & 0.30  & E & silica?   & 3, 7 \\  
9.45   & 0.19  & E &           & 3, 7 \\  
9.7   & 3    & A/E & amorphous silicate         & 1--3, 5--9 \\
9.8   & 0.17  & A/E & forsterite and enstatite   & 3, 7 \\  
9.8   & 0.4    & A & CH$_3$OH        & 2 \\
10.7 & 0.28  & A/E & enstatite   & 3, 7 \\  
11.2  & 1.7  & A/E & SiC              & 4, 5 \\ 
11.2  & 0.27  & A/E &  PAH  (H--C--H)   & 1--5, 7, 8 \\ 
11.4 & 0.48  & E & forsterite, diopside?   & 3, 7 \\  
12.0  & 0.47  & A/E &  H$_2$O         &  2 \\
12.7  &       & E &  PAH  (H--C--H)     &  2, 5, 7, 8 \\ 
13.0 & 0.6    &  E   & spinel            & 3 \\ 
13.3 &     &  E   & PAH?            & 7 \\ 
13.5 & 0.25 & E & ?    & 3 \\ 
13.8 & 0.20 & E & enstatite?    & 3 \\ 
13.6 &     &  E   & PAH?            & 7, 8? \\ 
14.2   & 0.28 &  E & enstatite?  & 3 \\ 
14.97  & 0.02 & A & CO$_2$            & 2         \\
15.2  &  0.06    & A &  CO$_2$      &  2, 9 \\
15.2   & 0.26 & A/E & enstatite  & 3, 4, 7 \\ 
15.8 &     &  E   & PAH?            & 7, 8 \\ 
15.9   & 0.43 & A/E & silica?  & 3, 4, 7 \\ 
16.2   & 0.16 & A/E  & crystalline forsterite & 3, 4  \\
16.4 &     &  E   & PAH?            & 7 \\ 
16.0 &     &  E   & spinel            & 3 \\ 
16.9   & 0.57 & A/E  &  ?                 & 3, 4, 7 \\ 
17.5   & 0.18 & E  & enstatite  &  3, 7\\ 
18.0   & 0.48 & A/E  & forsterite and enstatite  & 3--5, 7 \\
18.5  & 3  & A/E & amorphous silicate           & 1--3, 5--9 \\
18.9   & 0.62 & A/E  & forsterite?  & 3--5, 7 \\ 
19.5   & 0.40 & A/E & cryst. forsterite and enstatite & 3--5, 7 \\
20.7   & 0.31 & A/E & silica?, diopside?  & 3--5, 7 \\ 
21    & 5   & E & ?               & 4, 5\\ 
21.5   & 0.35 & E & ?  & 3--5, 7 \\ 
22.4   & 0.28 & E &       ?             & 3--5, 7 \\ 
23.0  & 0.48 & E & crystalline enstatite  & 3--5, 7 \\
23.7   & 0.79 & E & crystalline forsterite  & 3--5, 7 \\ 
23.89  & 0.18 & E & ? & 3, 5, 7 \\ 
24.5   & 0.42  & E & crystalline enstatite + ? & 3--5, 7 \\ 
25.0   & 0.32 & A/E  &  diopside?       & 3--5, 7 \\ 
26.1   & 0.57 & A/E & forsterite + silica? & 3--5, 7 \\ 
26.8   & 0.37 & A/E & ? & 3--5, 7 \\ 
27.6  & 0.49 & E & crystalline forsterite & 3--5, 7 \\ 
28.2  & 0.42 & E & crystalline enstatite  & 3--5, 7 \\ 
28.8  & 0.24 & E & ?  & 3, 5, 7 \\ 
29.6  & 0.89 & E & diopside?  & 3--5, 7 \\ 
30    & 20   & E & MgS               & 4, 5\\ 
30.6  & 0.32 & E &  ?         &  3--5, 7 \\ 
31.2  & 0.24 & E &  forsterite?         &  3, 4, 7 \\ 
32.0 &  0.5   &  E   & spinel            & 3 \\ 
32.2   & 0.46 & E & diopside?   & 3--5, 7 \\ 
32.8   & 0.60 & E & ?   & 3--5, 7 \\ 
33.6   & 0.70 & E & crystalline forsterite  & 3--5, 7 \\ 
34.1   & 0.12 & E & crystalline enstatite + diopside?  &3--5, 7 \\ 
34.9   & 1.36 & E & clino-enstatite?  & 3--5, 7 \\ 
35.9   & 0.53 & E & orto-enstatite?  & 3--5, 7 \\ 
36.5   & 0.39 & E & crystalline forsterite + ? & 3--5, 7 \\ 
38.1   & 0.57  & E & ? & 3 \\
39.8   & 0.74 & E & diopside?   & 3--5, 7 \\ 
40.5   & 0.93  & E & crystalline enstatite  & 3--5, 7 \\ 
41.8   & 0.72 & E &    ?               & 3 \\
43.0   &  & E & clino-enstatite   & 3--5, 7 \\
43--45   &     & A/E &  H$_2$O         &  2, 3--5, 7 \\ 
43.8  & 0.78 & E & orto-enstatite  & 3--5, 7 \\ 
44.7  & 0.58 & E & clino-enstatite, diopside?  & 3--5, 7 \\ 
47.7   & 0.97 & E &     FeSi?, a silicate    & 3--5, 7 \\ 
48.8   & 0.61 & E &     a silicate    & 3--5, 7 \\ 
52.9  & 3.11 & E & crystalline H$_2$O  & 3--5, 7 \\ 
62.    & 20 & E & crystalline H$_2$O      & 3--5, 7 \\ 
65.    &    & E & enstatite?, diopside?   & 3--5, 7 \\ 
69.0  & 0.63 & E & crystalline forsterite & 3--5, 7 \\ 
91.  &      & E & ? & 4, 5, 7 \\ 
\end{longtable}
\protect\vspace*{-0.3cm}
$^\ast\Delta\lambda$=FWHM (Full Width Half Max);
$^{\ast\ast}$chemical bond responsible for given feature is shown in parentheses;
A/E -- absorption/emission;
hydrocarbons:  \mbox{--CH$_2$--,} --CH$_3$ groups in aliphatic solids;
phyllosilicates: e.g., serpentine (Mg$_3$Si$_2$O$_{5}$[OH]$_4$)
or talc (Mg$_3$Si$_4$O$_{10}$[OH]$_2$);
metal carbonils:  e.g., Fe(CO)$_4$;
spinel:  MgAl$_2$O$_4$;
PAH -- polycyclic aromatic hydrocarbons;
HAC -- hydrogenated  amorphous carbon;
1 -- diffuse interstellar clouds;
2 -- molecular clouds and/or  HII compact regions ;
3 -- O-rich stars \mbox{\rm O/C $>$ 1};
4 -- C-rich stars \mbox{\rm C/O $>$ 1};
5 -- planetary nebulae and  HII regions;
6 -- eruptive variables;
7 -- other galactic sources;
8 -- galactic nuclei;
9 -- comets}

\vspace*{0.5cm}

Linear polarization of the IR radiation has a rather complicated wavelength
dependence and complex pattern. In the near-IR, the polarization is due
to scattering while the polarization vectors mark the position of
the illuminating source(s) \cite{wga00}.
The scattering efficiency of dust grains
sharply drops with wavelength and the polarization mechanism is
switched from scattering to dichroic extinction and thermal polarized
emission. The latter dominates at the far-IR and submillimeter
wavelength range.
Polarization due to dichroic extinction and thermal emission is attributed
to the spinning non-spherical dust grains aligned by magnetic fields.
The direction of the observed polarization is {\it parallel}
(dichroic extinction) or {\it perpendicular} (thermal emission)
to the magnetic field
as projected onto the plane of the sky.
One special problem is the observed polarization profiles of dust
features where both absorption and emission processes can occur
in a single observational beam (see Ref.~\cite{setal00} for discussion).

\subsection{Interpretation}

The modeling of the IR radiation is usually based on some
radiative transfer calculations. Even if an object is optically
thin at a given wavelength, the determination of the dust temperature
requires the consideration of the UV--visual radiation where, as a rule,
absorption by dust is large and the object's
optical thickness is significant.

The IR flux at wavelength $\lambda$ emerging from an optically
thin medium is proportional to the total number of dust grains 
in the medium $N$, the Planck function which depends on the particle
temperature $T_{\rm d}$, and the emission cross-section
${C}_{\rm em}(\lambda)$:
\be
F_{\rm IR}(\lambda) = N \ \frac{{C}_{\rm em}(\lambda)}{D^2}
B_{\lambda}(T_{\rm d}) =
 N \ \frac{\pi r_{\rm s} Q_{\rm abs}(\lambda)}{D^2} B_{\lambda}(T_{\rm d}),
\label{f}
\ee
where $D$ is the distance to the object.
The right-hand side of Eq.~(\ref{f}) is written with the assumption that the grains
are spheres of the same radius $r_{\rm s}$.

\setcounter{figure}{13}
\begin{figure}
 \epsfig{file=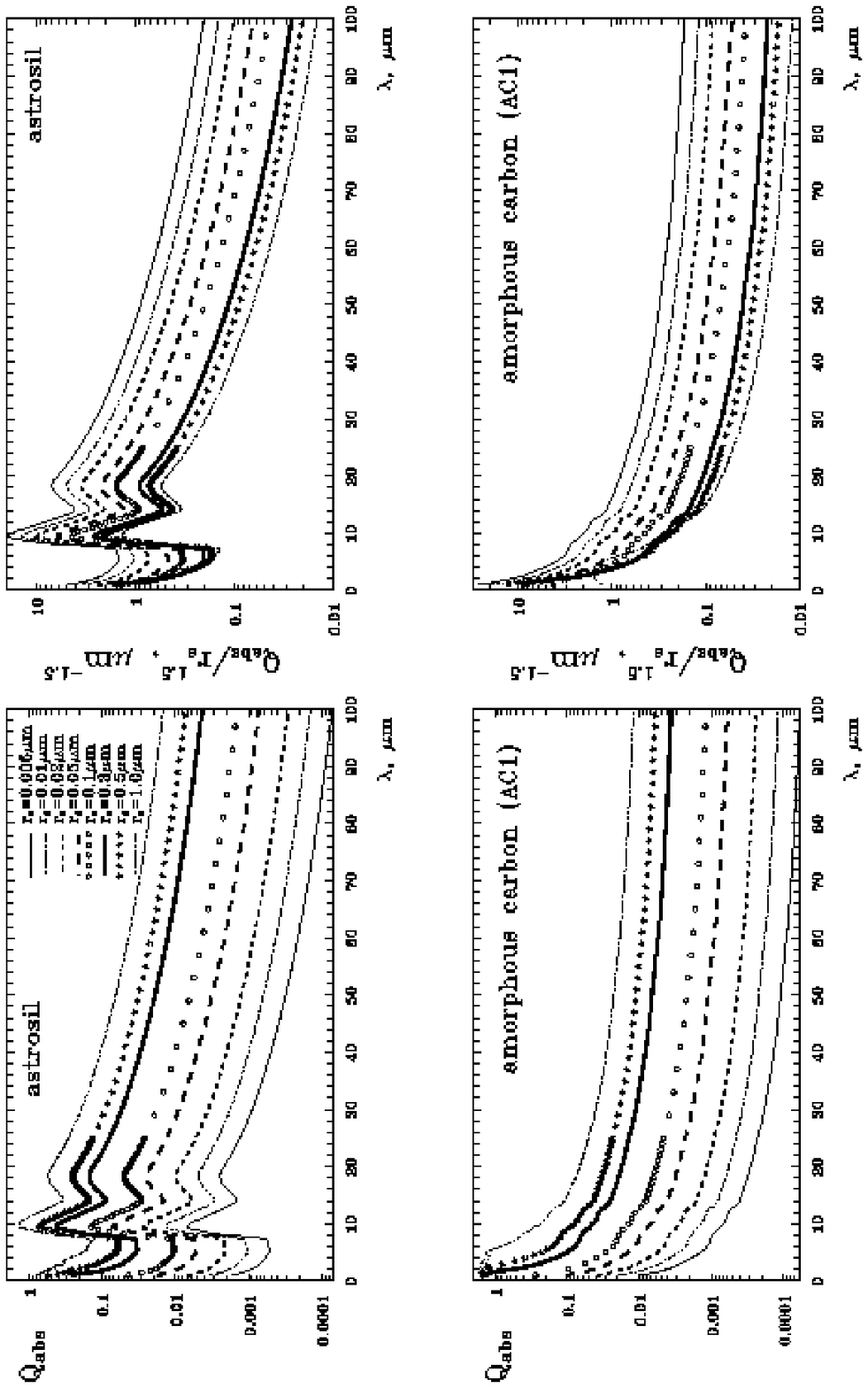,height=18cm,width=9.5cm}
\begin{turn}{90}
{\footnotesize
\setlength{\textheight}{0.5\baselineskip}
\protect\vspace{0.5cm}
{\it Figure \thefigure}. \,\,\,\,
Wavelength dependence of the absorption efficiency factors
for homogeneous spherical particles of different sizes
  }%
\end{turn}
\begin{turn}{90}
{\footnotesize
\setlength{\textheight}{0.5\baselineskip}
\protect\vspace{-0.5cm}
consisting
of astronomical silicate and amorphous carbon.
The right panels allow one to estimate the contribution
of particles into
  }%
\end{turn}
\begin{turn}{90}
{\footnotesize
\setlength{\textheight}{0.5\baselineskip}
\protect\vspace{-0.5cm}
thermal radiation
if the size distribution is like MRN.
}%
\end{turn}
\label{a_w}
\end{figure}
The wavelength dependence of the absorption efficiency factors
$Q_{\rm abs}(\lambda)$ is shown in Fig.~13 (left panels)
for spheres consisting of astrosil and amorphous carbon.
At the IR wavelengths, the factors $Q_{\rm abs}$ usually increase with
grain size and are larger for carbonaceous and metallic
particles in comparison with those of silicates and ices.
However,  the contribution of the particles of different sizes
into thermal radiation reverses if we take into account their
size distribution like MRN (Fig.~13, right panels).

\begin{figure}[htb]\bc
\resizebox{\hsize}{!}{\includegraphics{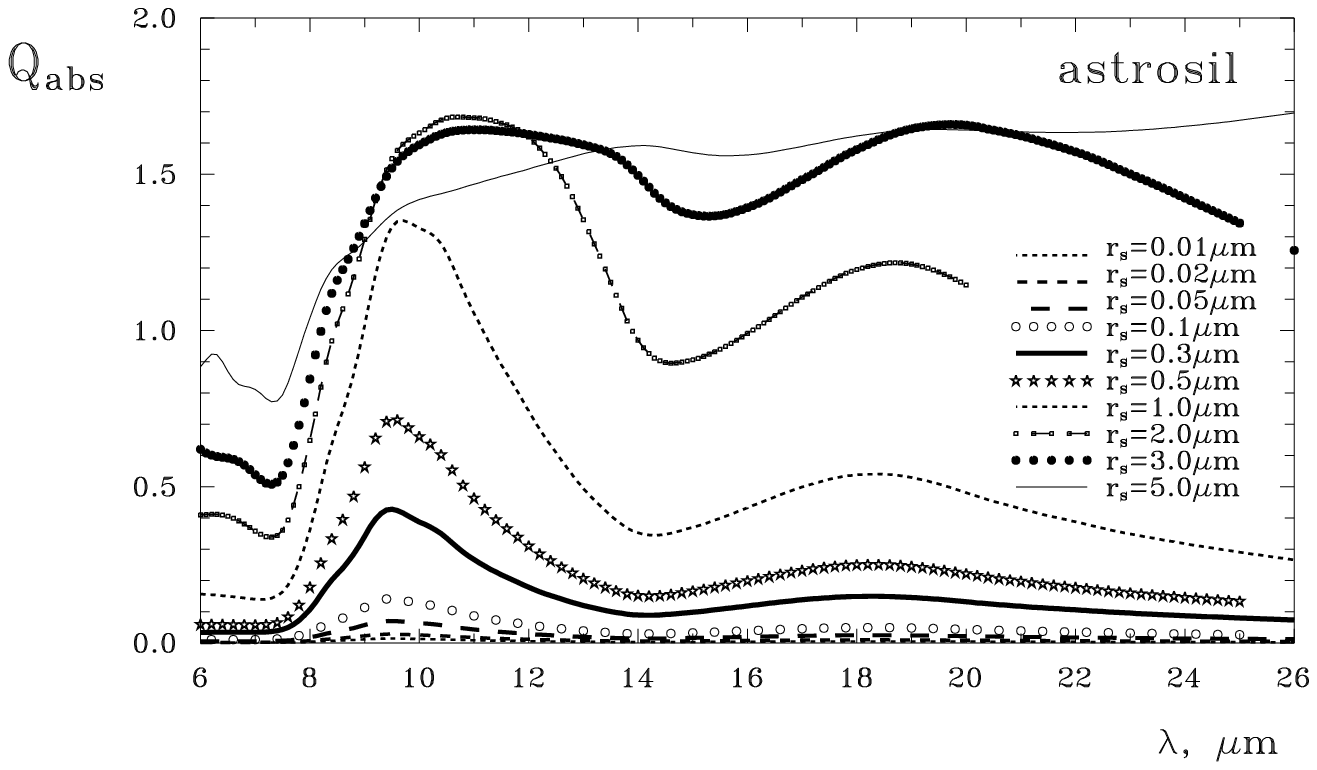}}
\caption{Wavelength dependence  of the absorption efficiency factors
for homogeneous spherical particles of astronomical silicate.
The effect of variation of particle
size is illustrated.
}\label{a_w10}
\ec
\end{figure}

It is important to note that the IR spectrum of carbonaceous
and metallic particles is almost
featureless (see also Fig.~\ref{f_riw} and Table~\ref{dust_f}). At the same time,
as follows from Fig.~\ref{a_w10}, the shape of dust features and even the
presence of them can tell us about the size of dust grains. If we consider
spherical grains of astrosil,
the 10-$\mkm$ and  18-$\mkm$  features disappear if the grain radius
exceeds $r_{\rm s} \ga 2-3 \,\mkm$ (Fig.~\ref{a_w10}).

The quantities $T_{\rm d}$ and
${C}_{\rm em}(\lambda)$  in Eq.~(\ref{f}) depend on the particle shape.
The shape dependence on temperature of interstellar/\cs dust grains
were analyzed in \cite{vs00}, \cite{vsh99}.
It was found that the temperature of
non-spherical (spheroidal) grains with aspect ratios $a/b \la 2$
deviates from that of spheres by less than 10\,\%.
More elongated or flattened particles are usually
cooler than spheres of the same mass and,
in some cases, the temperatures may differ by even about a factor of 2.


The causes of temperature differences between dust grains with different
characteristics can be clearly established with the aid
the graphical method proposed by Greenberg~\cite{g71}.
The particles are assumed to be in an isotropic radiation field and
their temperature can then be determined by solving the following
equation of thermal balance ($W$ is the dilution factor):
\be
W\,\int^\infty_0 {C}_{\rm abs}(\lambda) \ 4 \pi B_{\lambda}(T_{\star}) \,{\rm d}\lambda =
\int^\infty_0 {C}_{\rm em}(\lambda) \ \pi B_{\lambda}(T_{\rm d}) \,{\rm d}\lambda\,. \label{eq1a}
\ee
If the spheroidal particles are randomly oriented in space
(3D orientation), then numerical estimates show that
\be
{C}^{\rm 3D}_{\rm abs} \approx \frac{{C}_{\rm em}}{4}.
\ee
The integrals on the left- and right-hand sides of Eq.(\ref{eq1a}) then depend
on the temperature $T$ alone (for given particle chemical composition,
size, and shape).

\begin{figure}[htb]\bc
\resizebox{\hsize}{!}{\includegraphics{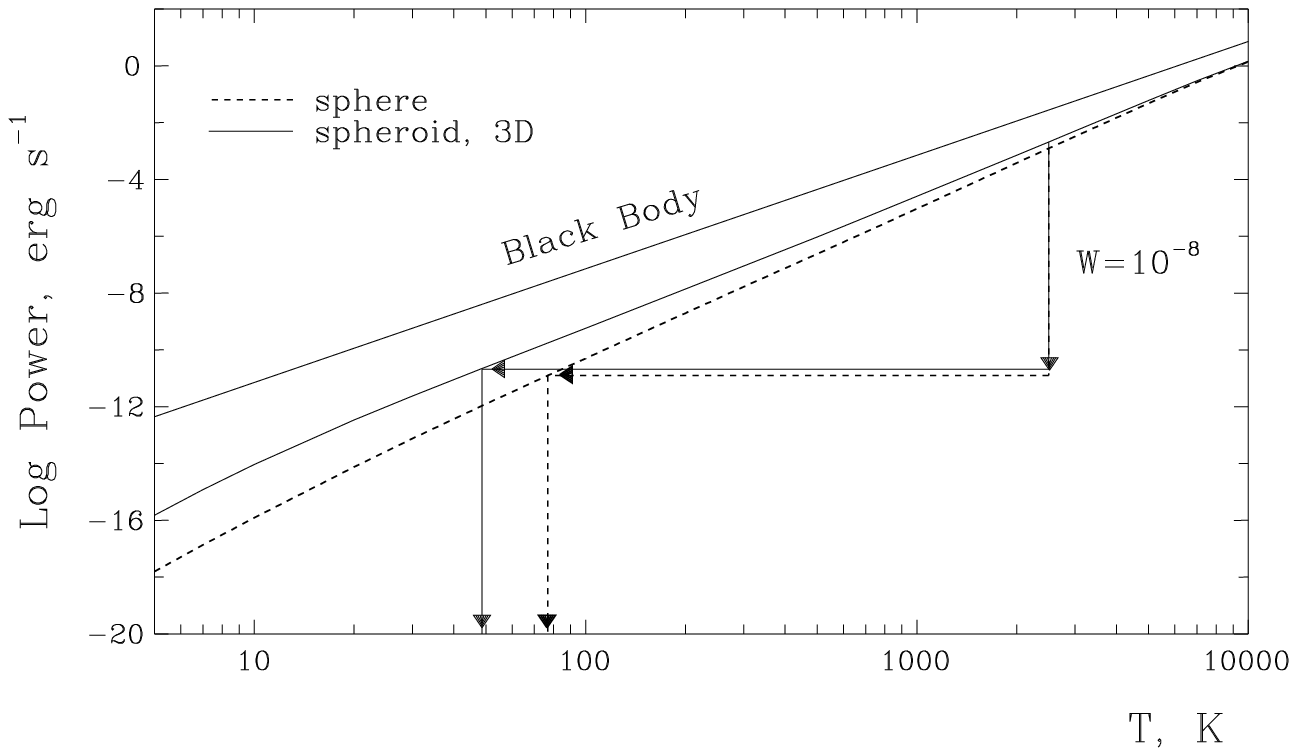}}
\caption{
The power absorbed (emitted) by dust particles
with $r_V=0.01\,\mkm$ in an isotropic radiation field.
The straight line corresponds to a blackbody.
The curves refer to spherical and prolate spheroidal
($a/b=10$, 3D orientation) cellulose (cel1000) particles.
A graphical method of determining the grain temperature
is shown for $T_{\star} = 2500$~K and W$ = 10^{-8}$. In this case,
the temperatures of spherical and spheroidal particles
and a blackbody are respectively,
$T_{\rm d}({\rm sphere})=76.8$~K,
$T^{\rm 3D}_{\rm d}=48.7$~K, and
$T_{\rm d}^{\rm BB}= 2500 (10^{-8})^{1/4}= 25$~K 
(after \protect \cite{vs00}).
}
\label{f1}
\ec\end{figure}

Figure~\ref{f1}  shows the power for the absorbed (emitted) radiation (in erg s$^{-1}$)
for spheres and spheroids of the same volume ($r_V = 0.01\,\mkm$)
composed of cellulose\footnote{This is the
carbon material pyrolized at 1000$\degr$\,C
(cel1000, \cite{j98}).}.
The method of temperature determination
is indicated by arrows: from the point on the curve corresponding
to the stellar temperature ($T_{\star}$ = 2500 K), we drop a perpendicular whose
length is determined by the radiation dilution factor and then find the
point of intersection of the horizontal segment with the same curve
for the power. This point gives the dust grain temperature $T_{\rm d}$ determined
from the equation of thermal balance (\ref{eq1a}).
It follows from Fig.~\ref{f1} that different temperatures of spheres and
spheroids result from different particle emissivities at low $T$.

\begin{figure}\bc
\resizebox{\hsize}{!}{\includegraphics{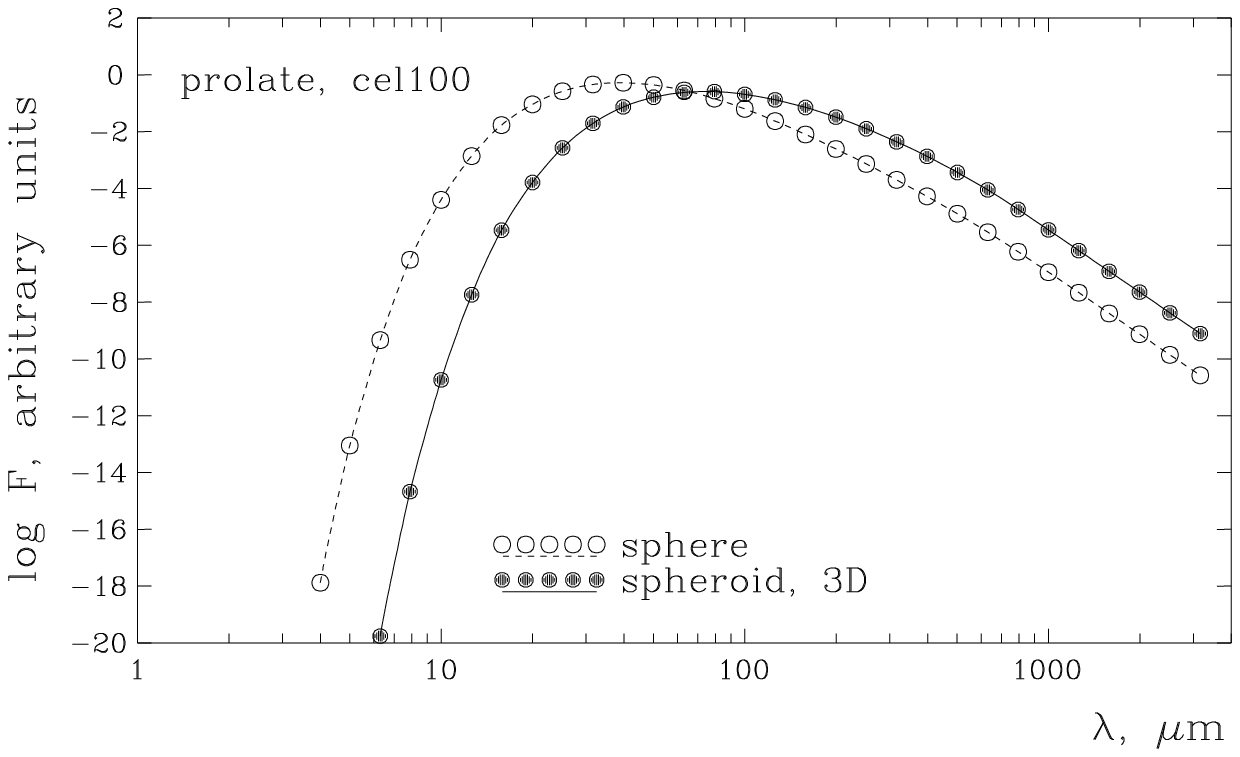}}
\caption{The normalized fluxes emerging from a medium containing
the same mass of spherical and prolate spheroidal
($a/b = 10$, 3D orientation) cellulose (cel1000) particles;
$r_V=0.01\,\mkm$, $T_{\star}$ = 2500 K, and $W = 10^{-8}$.
The particle temperatures are $T_{\rm d}({\rm sphere})=$ 59.2~{\rm K}
and $T^{\rm 3D}_{\rm d}=$ 36.0~K 
(after \protect \cite{vs00}).
}\label{t}
\ec
\end{figure}

If we assume the chemical composition and sizes of all particles
in the medium to be the same and if we change only the grain shape,
then, when spheres are replaced by non-spherical particles,
the position of the maximum in the spectrum of thermal radiation
is shifted longward, because the temperature of the latter is higher (Fig.~\ref{t}).

The dust mass in an object $M_{\rm d}$  is determined from the observed
flux $F_{\rm IR}(\lambda)$ and the mass absorption coefficient of
a grain material $\kappa(\lambda)$
\be
M_{\rm d} =  \frac{F_{\rm IR}(\lambda) D^2}{\kappa(\lambda) B_\lambda(T_{\rm d})}
=\frac{\rho_{\rm d} V F_{\rm IR}(\lambda) D^2}
{C_{\rm abs}(\lambda) B_\lambda(T_{\rm d})},
        \label{m}
\ee
where $V$ is  the particle volume and $\rho_{\rm d}$ the material density.
Extensive study of the dependencies of the mass absorption coefficients
on material properties and grain shape is given in \cite{ks94},
\cite{oh94} and summarized in \cite{h96} where,
in particular, it is shown that the opacities at 1~mm are considerably
larger for non-spherical particles in comparison with spheres.

If $M_{\rm d}$ is determined from the observed
millimeter flux, then the Ray\-leigh--Jeans approximation can be used for
the Planck function. For the same flux $F_{\rm IR}(\lambda)$, the dust mass
depends on the particle emission cross sections and temperatures,
while the ratio
\be
\frac{M^{\rm sphere}_{\rm d}}{M^{\rm spheroid}_{\rm d}} =
 \ \frac{{C}^{\rm spheroid}_{\rm em}(\lambda)}
        {C^{\rm sphere}_{\rm em}(\lambda)}
 \ \frac{T_{\rm d}({\rm spheroid})}
        {T_{\rm d}({\rm sphere})}
        \label{mm}
\ee
shows the extent to which the values of $M_{\rm d}$ differ when changing the
grain shape.
The dust mass in galaxies and molecular clouds is commonly
estimated from 1.3-mm observations \cite{setal99}. At
this wavelength, the ratio of the cross-sections for cellulose particles
with $a/b = 10$ is $\sim$~50, and the temperature
ratio is $36.0/59.2 = 0.61$, which, according to Eq.~(\ref{mm}),
gives approximately a factor of
30 larger dust mass if the particles are assumed to be spheres rather
than spheroids. In other, not so extreme, cases, the dust mass
overestimate is much smaller. For example, for prolate amorphous
carbon particles,
$M^{\rm sphere}_{\rm d} / M^{\rm spheroid}_{\rm d} \approx 1.2, 2.3$, and 6.4,
for $a/b =$ 2, 4, and 10, respectively.

Analysis of the polarimetric observations
is based on models where dust properties and temperature,
as well as the object's opacity, are varied. The main purpose
of such modeling is to identify the magnetic field
configurations (see \cite{aitetal02}, \cite{hetal00} for discussion).
Unfortunately, the models usually contain many parameters and are not
unique.

\section{What do we really know about the cosmic dust?}

Our current knowledge of cosmic dust
and the possibility of extracting information
about it from existing and future observations can be summarized
as follows.

The study of the \is extinction and polarization, together with
the constraints from the cosmic abundances, allows one to estimate
the chemical composition, size and shape of dust grains.
These three characteristics cannot be found separately but only
as a combination. At the same time, it may be argued
that the orientation angle of the \is linear polarization 
decisively provides the projected direction of alignment, i.e.
the projected direction of the \is magnetic field B$_{\bot}$.

Chemical composition, size and shape of dust grains
can be estimated approximately from the observed intensity, colors and
polarization of the scattered light. In this case, the
 polarization vectors may mark the position
of illuminating source(s).

By investigating the thermal emission of dust,
one can evaluate with confidence grain temperature,
from observations in continuum, and
com\-po\-si\-ti\-on,\footnote{or more exactly chemical bonds in solids}
from observations of dust features.
Submillimeter polarization again shows
the projected direction of the \is magnetic field B$_{\bot}$.
The particles' size and shape also can be estimated approximately.

The current state of different components in the
modeling  of dusty objects can be summed up as
a rather well developed set of
optical constants, light scattering theories and
radiative transfer methods, which allow one to produce
models of very complicated objects.
However, the last stage of any modeling effort is a comparison with observations
where we are restricted by the Stokes vector of incoming
radiation $(I,Q,U,V)$.
{\it But}: the principle of optical equivalency
does work! This means that {\it all} models  are ambiguous
to some degree.

\acknowledgements{The author is thankful to V.B.~Il'in
and D.A.~Semenov for
valuable comments and to T.V.~Zinov'eva for the assistance with compilation
of Table~\ref{dust_f}.
This research was partly supported by the INTAS grant (Open Call 99/652)
and by grant 00-15-96607 of the President of the Russian Federation  
for leading scientific schools.}


{}

\end{document}